\documentclass[aps,prl,twocolumn,showpacs,superscriptaddress]{revtex4-2}
\usepackage{latexsym}
\usepackage{amssymb}
\usepackage{graphicx}
\usepackage{amsmath}
\usepackage{tikz}
\usepackage{bm}
\usepackage[colorlinks,linkcolor=red,citecolor=blue,urlcolor=blue]{hyperref}
\usepackage{verbatim}
\usepackage{bbding}
\usepackage{mathrsfs}
\usepackage{extarrows}
\usepackage{comment}
\usepackage{mathtools,slashed}
\usepackage{soul}
\usepackage[toc,page]{appendix}
\usepackage[vcentermath]{youngtab}
\usepackage{multirow}
\usepackage[width=.75\textwidth]{caption}
\usepackage{atbegshi,picture}
\usepackage{lipsum}

\usepackage{xr}

\makeatletter

\renewcommand\@makecaption[2]{
	\par
	\vskip\abovecaptionskip
	\begingroup
	\small\rmfamily
	\begingroup
	\samepage
	\flushing
	\let\footnote\@footnotemark@gobble
	\@make@capt@title{#1}{#2}\par
	\endgroup
	\endgroup
	\vskip\belowcaptionskip
}
\makeatother

\AtBeginShipoutNext{\AtBeginShipoutUpperLeft{
		\put(\dimexpr\paperwidth-1cm\relax,-1.5cm){\makebox[0pt][r]}
}}
\makeatletter
\begin{document}	
\title{Fracton Topological Order at Finite Temperature}
	
\author{Xiaoyang Shen}
\thanks{These authors contributed equally to the work.}
\affiliation{Institute for Advanced Study, Tsinghua University, Beijing 100084, China}

\author{Zhengzhi Wu}
\thanks{These authors contributed equally to the work.}
\affiliation{Institute for Advanced Study, Tsinghua University, Beijing 100084, China}

\author{Linhao Li}         
\thanks{These authors contributed equally to the work.}
\affiliation{Institute for Solid State Physics, The University of Tokyo. Kashiwa, Chiba 277-8581, Japan}

\author{Zhehan Qin}
\affiliation{Institute for Advanced Study, Tsinghua University, Beijing 100084, China}

\author{Hong Yao}
\thanks{yaohong@tsinghua.edu.cn}
\affiliation{Institute for Advanced Study, Tsinghua University, Beijing 100084, China}

\date{\today}

\begin{abstract}
As new kinds of stabilizer code models, fracton models have been promising in realizing quantum memory or quantum hard drives. However, it has been shown that the fracton topological order of 3D fracton models   
occurs only at zero temperature. In this Letter, we show that 
higher dimensional fracton models 
can support a 
fracton topological order below a nonzero critical temperature $T_c$. Focusing on a typical 4D X-cube model, we show that there is a finite critical temperature $T_c$ by analyzing its 
free energy from duality. We also obtained the expectation value of the 't Hooft loops in the 4D X-cube model, which directly shows a confinement-deconfinement phase transition at finite temperature. This finite-temperature phase transition can be understood as spontaneously breaking the $\mathbb{Z}_2$ one-form subsystem symmetry. Moreover, we propose a new no-go theorem for finite-temperature quantum fracton topological order. 
\end{abstract}
	
\maketitle
\textit{Introduction.---}
Reliable quantum information in realistic quantum computers requires the 
ability of error correction \cite{lidar2013quantum,shor1996fault,PhysRevLett.77.793,PhysRevA.52.R2493}.
 A theoretical proposal is the self-correcting quantum memory (SCQM) \cite{dennis2002topological,terhal2015quantum,brown2016quantum,PhysRevA.54.1098,nussinov2008autocorrelations}, whose memory time are expected to be made arbitrarily long with the increase of the system size even at finite temperature \cite{PhysRevX.10.031041,PhysRevLett.111.200501,brown2016quantum}.
Promising candidates for SCQM are the stabilizer code models with topological order   \cite{1997PhDT.......232G,kitaev2003fault,2003PhRvL..90a6803W,dennis2002topological}. They have degenerate ground states which cannot be distinguished by any local operators, and are thus robust quantum memory at zero temperature \cite{kitaev2003fault,2010JMP....51i3512B,Alicki_2007}. However, topological order is required to exist at a 
finite temperature to be SCQM \cite{YOSHIDA20112566},
whose expectation values of Wilson and 't Hooft loops are both nonzero. But there are no-go theorems excluding 2D and 3D stabilizer code with scale and translation symmetries (STS models) to have finite temperature quantum topological order \cite{Bravyi_2009,YOSHIDA20112566}. These models include the 2D and 3D toric code models, and only the toric code model in 4D has finite temperature quantum topological order, thus a candidate for SCQM \cite{dennis2002topological}.
 What's more, if only one kind of operators have nonzero expectation value at finite temperature, we may define this order as finite-temperature classical topological order \cite{PhysRevLett.107.210501,castelnovo2007entanglement}, such as 3D toric code model, which can be viewed as candidates for self-correcting classical memory. Recently, there are proposals of SCQM using 2D symmetry enriched topological order on the edge of 3D systems, if 1-form symmetries are enforced in the Hamiltonian and the dynamics \cite{PhysRevX.10.031041,Stahl_2021}. 
Fracton models \cite{PhysRevLett.94.040402,PhysRevB.92.235136,vijay2016fracton,Haah2011Local,Nandkishore2019fracton,pretko2020fracton,PhysRevB.95.155133,pretko2020fracton,haah2013lattice,vijay2017isotropic,PhysRevB.95.245126,2021tanti} are new kinds of stabilizer code models proposed recently, which are beyond the above no-go theorems. The 3D X-cube model, which is a prototypical fracton model, has very different properties with conventional topological order. Its ground state degeneracy(GSD) on the torus grows exponentially with the linear system size, while that of conventional topological order is a constant and independent of the system size. This is deeply rooted in the foliated structure of the fracton models \cite{10.21468/SciPostPhys.6.4.043,PhysRevLett.126.101603,SHIRLEY2019167922,PhysRevB.102.115103}. Thus, this kind of models is expected to encode substantially more quantum information compared with conventional topologically ordered models \cite{vijay2016fracton,PhysRevLett.111.200501,Haah2011Local,2020}. 
It is then highly desired to search for fracton models which can be employed as SCQM. For now, studies of possible finite-temperature fracton topological order (FTFTO) mainly focus on 3D fracton models, including the 3D X-cube model and the Haah's code. Although 
the Haah code and Chamon's models are proposed to be partially self-correcting in finite system size \cite{Haah2011Local,PhysRevLett.111.200501,RevModPhys.88.045005,castelnovo2012topological}, their topological order are thermally fragile in the thermodynamic limit \cite{weinstein2019universality,weinstein2020absence,li2019finite}. 
It is then natural to ask the following question: whether there is an exactly solvable model with classical/quantum FTFTO in higher dimensions? Meanwhile, the interesting physics of the 3D X-cube model also motivates the study of higher dimensional generalizations. The anyon excitations, different from conventional topological order, have restricted motion directions without the input of energy. For example, the fracton excitations are totally immobile, which is due to the dipole conservation and captured by their exotic effective tensor gauge theories \cite{PhysRevLett.124.050402,PhysRevB.97.235112,QI2021168360,ma2020fractonic,PhysRevLett.126.101603,PhysRevB.98.115134,SLAGLE2019167910,PhysRevB.98.035111,slagle2017quantum,10.21468/SciPostPhys.10.2.027,Seiberg:2020cxy,Seiberg:2020wsg,PhysRevB.95.115139,PhysRevB.96.035119}.

In this Letter, we investigate a series of higher dimensional fracton models with classical FTFTO.  We first focus on a 4D generalization of X-cube model with string-like excitations in one sector, which is similar to the 3D toric code model \cite{Castelnovo20083TC}. As a result, this part of partition function is regular near zero temperature which implies there is no phase transition at zero temperature in this sector. We show this part of partition function is partially dual to that of 3D toric code model. The other partition function sector is dual to Ising chain with zero temperature phase transition. These results together imply a classical FTFTO. Further, we show the physics can be understood by the generalized Elitzur’s theorem which prohibits the spontaneous symmetry breaking (SSB) of higher-form subsystem symmetry. The above discussions can be generalized to a new no-go theorem excluding quantum FTFTO in fracton models with one-form subsystem symmetry. 

\begin{figure}[t]
    \centering 
    \includegraphics[width=8.9cm]{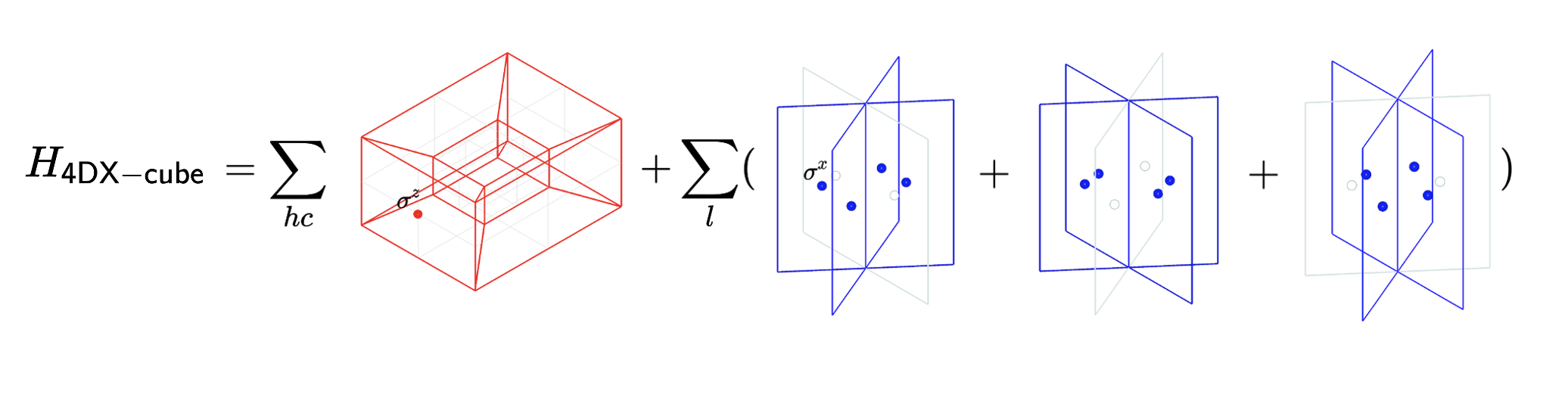} 
    \caption{The schematic representation of the 4D X-cube model in a pictorial language. On each plaquette there is a spin-1/2 degree of freedom. The red part are the hypercube terms, each of which is a product of $\sigma^z$'s on the 24 plaquettes of a hypercube, and the blue part are the link terms, each of which is a produce of $\sigma^x$'s on the 4 plaquettes sharing the link $l$.} 
    \label{type2}
\end{figure}

\textit{The 4D X-cube model.---}The 4D X-cube model on a cubic lattice with length $L$ is defined as:
\begin{equation}\label{X-cube}
   H = -J_A\sum_{hc} \mathcal{A}_{hc} - J_B\sum_{l\mu} \mathcal{B}_l^\mu,
\end{equation}
where $\mathcal{A}_{hc}$ and $\mathcal{B}_{l}^\mu$ are referred to as the hypercube and link terms, respectively. The index $l$ labels link \footnote{For a specific term, we use direction index to label the link.} and $\mu$ represents one direction perpendicular to link $l$, where $\mu=x,y,z,w$ ($w$ labeling the fourth direction in 4D). Both the hypercube and link terms are tensor product of Pauli matrices living on plaquettes of the hypercubic lattice:
\begin{equation}
  \mathcal{A}_{hc}=\prod_{i \in hc} \sigma_{i}^{z}, \quad \mathcal{B}_{l}^\mu=\prod_{i \in \{\bar S^\mu_l\}} \sigma_{i}^{x},
\end{equation}
where the set $\{\bar S^\mu_l\}$ consists of four plaquettes which share the link $l$ and are perpendicular to the surface $S_l^{\mu}$, where $S_l^{\mu}$ is the plaquette expanded by the directions $l$ and $\mu$. It is clear that every term in the Hamiltonian commutes with each other: 
\begin{equation}
   [\mathcal{A}_{hc_1}, \mathcal{A}_{hc_2}] = 0,\quad [\mathcal{B}_{l_1}^{\mu_1},\mathcal{B}_{l_2}^{\mu_2}] = 0,\quad [\mathcal{A}_{hc},\mathcal{B}_{l}^\mu] = 0.
\end{equation}
Consequently, the 4D X-cube model defined in Eq. \eqref{X-cube} is exactly solvable. Hereafter we set $J_A=J_B=1$ as energy unit and our results obtained below apply to generic values of $J_A$ and $J_B$. 

\textit{The effective field theory.---} Effective field theory plays an important role in understanding the physics of the lattice models with (fracton) topological order \cite{slagle2017quantum} and 'gauge structure' \footnote{There are infinite generators for gauge transformations.}, such as the 3D X-cube model, 3D toric code model, etc.  All the eigenstates and gauge invariant operators of the lattice model can be constructed from the effective field theory. The effective field theory here is especially useful to construct gauge invariant Wilson/'t Hooft operators directly without referring to the complicated four dimensional spatial geometry. The nonzero expectation value of non-contractible Wilson and 't Hooft operators is required for the X-cube phase. The canonical coordiante and momenta of the field theory are related to the lattice operators as:
\begin{equation}
\begin{aligned}
&\hat{Z}_{i}(t) \sim \exp \left(i \int_{S}  Z_{\alpha\beta}(\boldsymbol{x}, t)\right)=\exp \left(i \int_{S}   A_{\gamma \delta}|\epsilon^{0\alpha\beta\gamma\delta}|\right),\\
&\hat{X}_{i}(t) \sim \exp \left(i \int_{\perp S} X_{\alpha\beta}(\boldsymbol{x}, t)\right)=\exp \left(i \int_{\perp S} B_{\gamma\delta} |\epsilon^{0\alpha\beta\gamma\delta}|\right),
\end{aligned}
\end{equation}
where the integration regime $S$ is the plaquette of the lattice operators and $\perp S$ is its Poincare dual. The label $\alpha\beta$ are the spatial directions of the plaquette. The hypercube terms and link terms are conserved charges and generators of the gauge transformation of the field theory, which means the Lagrangian density is:
\begin{eqnarray}\label{BF}
    &&\mathcal{L}_{\textsf{X-cube}} = \frac{1}{\pi}A_{\alpha\beta}\partial_{0}B_{\alpha\beta}+B_{0}(\frac{1}{\pi}\partial_{\alpha}\partial_{\beta}A_{\alpha\beta}-i^0)\nonumber\\
    &&+A_{0;\alpha\beta}(\frac{1}{\pi}\epsilon^{0\alpha\beta\gamma\delta}\partial_{\gamma}B_{\beta\delta}-j^{0;\alpha\beta})-A_{\alpha\beta}j^{\alpha\beta}-B_{\alpha\beta} i^{\alpha\beta},~~~
    \end{eqnarray}
where $A$ and $B$ are rank-2 tensor gauge fields, and the time and spatial components of $i,j$ correspond to the excitation density and current configuration.  The time components of the gauge fields are denoted as $A_{0;\alpha\beta}$ and $B_0$, and the spatial components are denoted as $A_{\alpha\beta}$ and $B_{\alpha\beta}$. The gague invariant contractible(noncontractible) Wilson/'t Hooft operators can be constructed from the field theory and are illustrated in Fig.\ref{loop} and  Fig.\ref{topological} respectively. The detailed construction is left in the supplementary material.
\begin{figure}[t]
    \centering 
    \includegraphics[width=4cm]{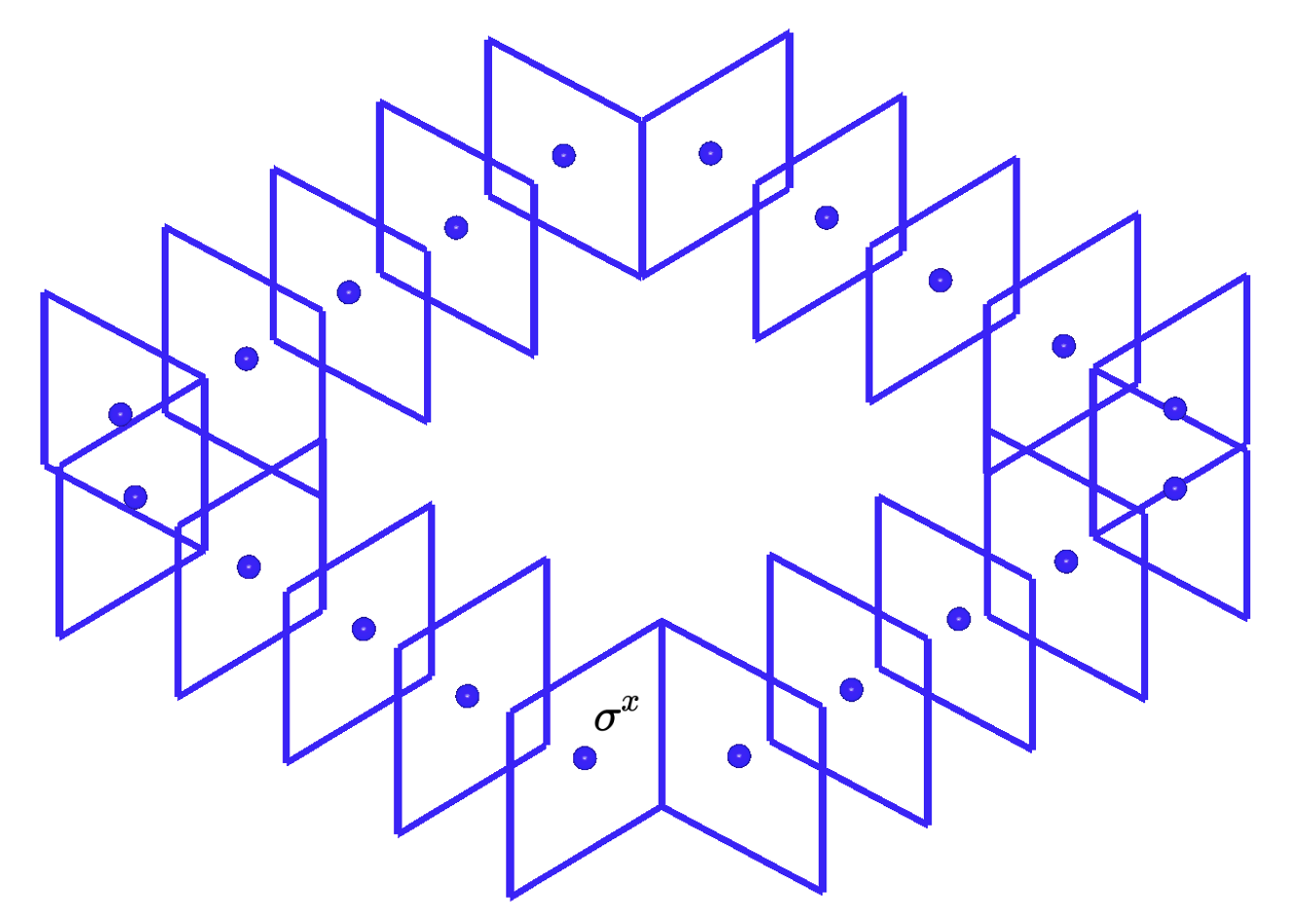} 
    \caption{The schematic representation of a contractible ’t Hooft loop operator $T_\gamma=\prod_{r\in \gamma}\sigma^x(r)$, where $\gamma$ labels the loop formed by the centers of plaquettes along the closed contour.}
    \label{loop}
\end{figure}
\begin{figure}[t]
    \centering 
    \includegraphics[width=4cm]{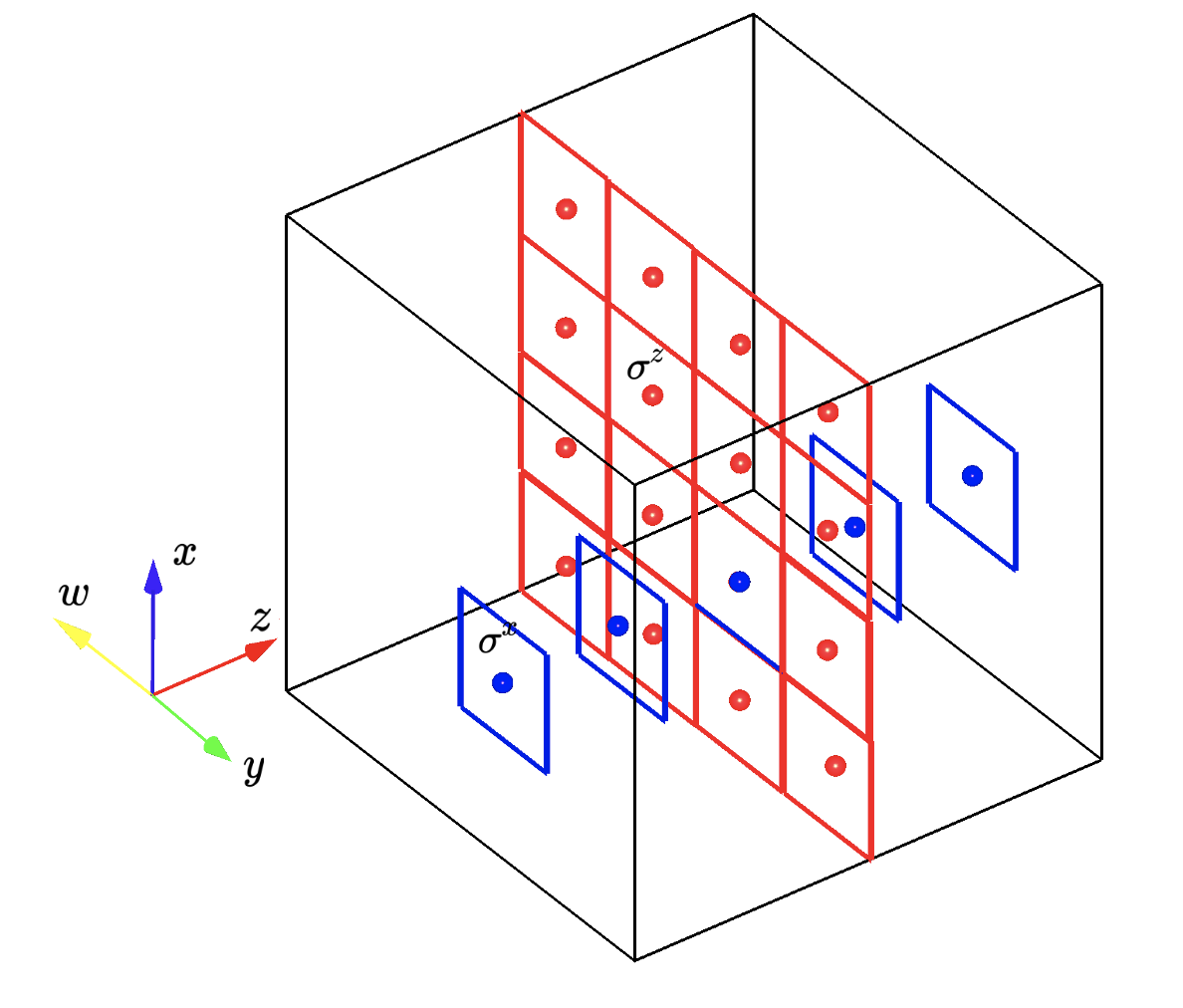} 
    \caption{The red and blue operators are non-contractible Wilson loop and 't Hooft loop operators respectively.}
    \label{topological}
\end{figure}


 
\textit{The partition function duality.---}For the partition function of the 4D X-cube model $\mathcal{Z}=\operatorname{Tr}(e^{-\beta H})$, the link part and hypercube part are decoupled due to the facts that $\mathcal{A}_{hc}$ and $\mathcal{B}_{l}^\mu$ are products of traceless matrices $\sigma^z,\sigma^x$ and the product of $\mathcal{A}_{hc}$ and $\mathcal{B}_{l}^\mu$ is thus traceless. Consequently, these two parts can be calculated independently: $\mathcal{Z}=\frac{1}{d_h}\mathcal{Z}_{\textsf{hypercube}}\mathcal{Z}_{\textsf{link}}$, where $d_h=\operatorname{Tr}[\mathbb{I}]$ is the dimension of the Hilbert space, $\mathcal{Z}_{\textsf{hypercube}}=\operatorname{Tr}(e^{\beta\sum_{hc} \mathcal{A}_{hc}})$, $\mathcal{Z}_{\textsf{link}}=\operatorname{Tr}(e^{\beta\sum_{l\mu} \mathcal{B}_l^\mu})$. Here we use open boundary condition (OBC) \cite{weinstein2020absence}\footnote{The periodic boundary condition(PBC) will give the same partition function as OBC. See the Supplemental Material for the proof.}. 

Let's first consider the hypercube part $\mathcal{Z}_{\textsf{hypercube}}=\operatorname{Tr}\left[\prod_{hc}\left(\mathbb{I} \cosh\beta+\mathcal{A}_{hc} \sinh\beta\right) \right]$ under OBC, where the only nonzero contribution is the multiplication of all the identity matrices in the series expansion. So we have $\mathcal{Z}_{\textsf{hypercube}} = d_h(\cosh\beta)^{L^4} \sim (e^{\beta}+e^{-\beta})^{L^4}$, where $L^4$ is the number of lattice sites with linear size $L$. 
Thus, $\mathcal{Z}_{\textsf{hypercube}}$ is dual to $L^4$ isolated spins in a magnetic field under OBC, which is in turn dual to an open Ising chain of length $L^4+1$. 

Next, we consider the link part $\mathcal{Z}_{\textsf{link}}=\operatorname{Tr}\prod_{l\mu}(\mathbb{I}\cosh\beta +\mathcal{B}_l^\mu\sinh\beta)$. Note that not all $\mathcal{B}_l^\mu$ are independent; namely there are local constraints for $\mathcal{B}_l^\mu$. The local constraints can be divided into two types. The first type (dubbed as type-I constraints) 
refers to 
the constraint that the product of the three link terms $\mathcal{B}_{l}^\mu$ sharing the same link $l$ equals to identity:
   $\prod_{\mu}\mathcal{B}_{l}^\mu = \mathbb{I}$,
which is a local constraint for each $l$. 
The type-I constraints render a contribution to the partition function:  $\mathcal{Z}_{\textsf{link(type-I)}}$, which is equivalent to $4(L-1)^4$ independent three-spin clusters $\mathcal{Z}^{4(L-1)^4}_{\textsf{3-spin Curie-Weiss}}$. 

\begin{figure}[t]
    \centering 
    \includegraphics[width=6cm]{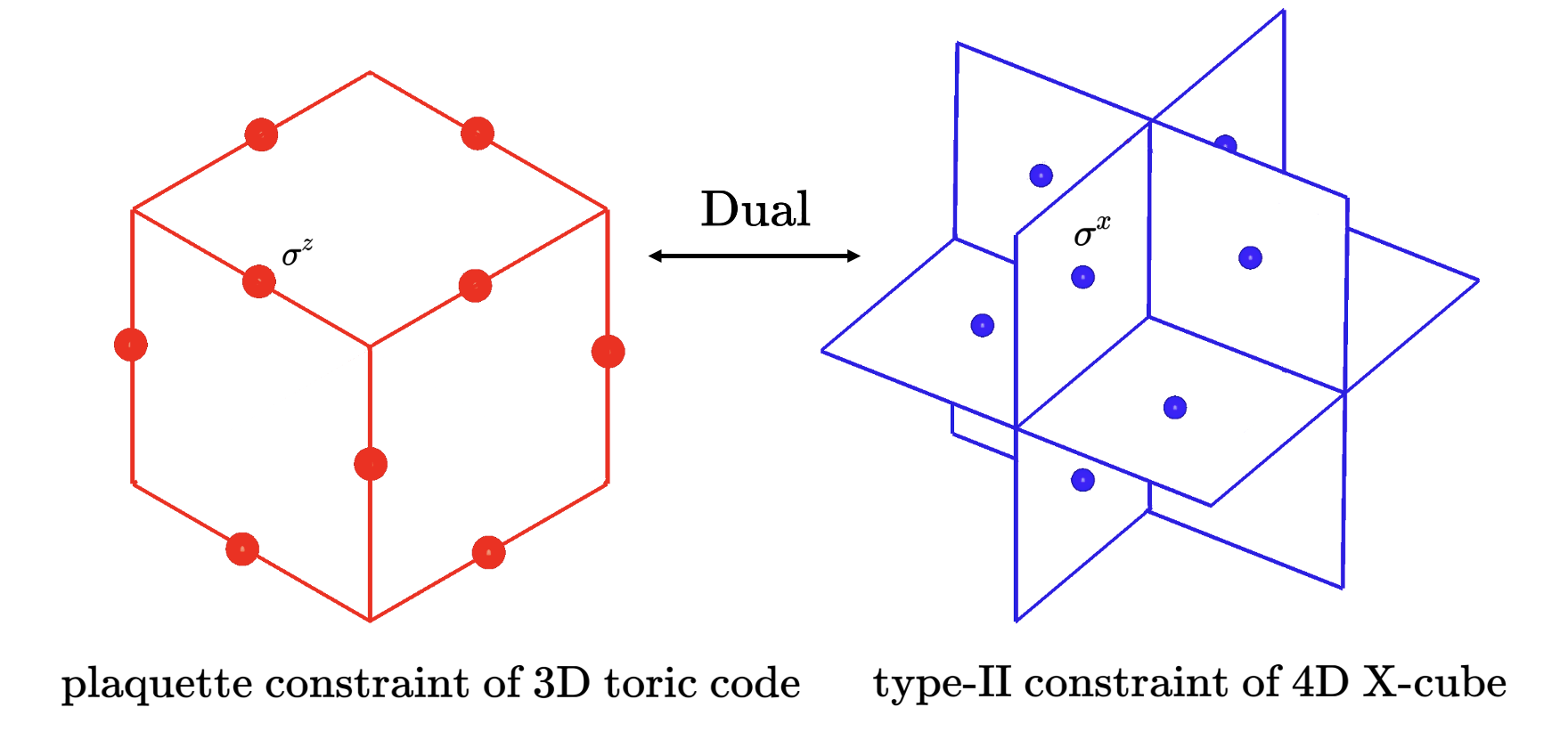} 
    \caption{An illustration of the duality between the type-II constraints in the 4D X-cube model and the plaquette constraints in the 3D toric code model.}
    \label{duality}
\end{figure}

The second type constraints (type-II constraints) have no counterparts in the 3D X-cube model. As shown in Fig. \ref{duality}, the product of six link terms sharing the same vertex  equal to identity: 
    $\prod_{l}\mathcal{B}^\mu_l \mathcal{B}^\mu_{\bar{l}} = \mathbb{I}$,
where link $l$ and $\bar{l}$ have opposite directions. Written explicitly, one example of type-II constraints is $\mathcal{B}_x^w \mathcal{B}_y^w \mathcal{B}_z^w \mathcal{B}_{\bar{x}}^w \mathcal{B}_{\bar{y}}^w \mathcal{B}_{\bar{z}}^w=\mathbb{I}$. 
The type-II contribution $\mathcal{Z}_{\textsf{link(type-II)}}$ is  dual to $ (\mathcal{Z}_{\mathcal{B}_p})^{4L} $, where $\mathcal{B}_p = \prod_{i\in \text{plaquette(p)}}\sigma^z_i$ is the plaquette term in 3D toric code model. This duality is rooted in the identification of the type-II constraints with the toric code constraints under the lattice duality:
$\prod_{p \in \text{cube}} \mathcal{B}_{p}=\mathbb{I}\label{cubeconstraint}$.
If we dual the link in the toric code model into the perpendicular surface, then the constraint on the plaquette terms in the 3D toric code model is dual to the type-II constraint, which is shown in the Fig.  \ref{duality}. The duality of the type-II contribution is: $ \mathcal{Z}_{\textsf{hypercube}}\times\mathcal{Z}_{\textsf{link(type-II)}}\sim(\mathcal{Z}_{\textsf{3D TC}})^{4L}/\mathcal{Z}_{\textsf{1D-Ising}}$. The detailed number counting of both types contribution is in the supplementary material (SM). Since the 3D toric code features a finite-temperature transition, we had proven exactly that the 4D X-cube model also has a finite-temperature transition if type-I and type-II constraints are totally independent. Nonetheless, type-I and type-II constraints can coexist locally; for instance the type-I constraint $\mathcal{B}^x_{w}\mathcal{B}^y_{w}\mathcal{B}^z_w=1$ and type-II constraint $\mathcal{B}^x_w\mathcal{B}^x_y\mathcal{B}^x_z\mathcal{B}^x_{\bar w}\mathcal{B}^x_{\bar y}\mathcal{B}^x_{\bar z}$ share the link term $\mathcal{B}^x_w$ such that $\mathcal{Z}_\textsf{link}$ cannot be simply written as $\mathcal{Z}_\textsf{link(type-I)}\mathcal{Z}_\textsf{link(type-II)}$. 
Although it is challenging to find an exact duality of the partion function, 
we believe that a finite-temperature phase transition should survive due to type-II constraints. Indeed, in the following sections we can prove the existence of a finite-temperature transition in the 4D X-cube model by using low/high temperature expansion and effective field theory analysis.


\textit{Low-temperature free energy expansion.---}In this section, we conduct the low-temperature free energy expansion \cite{kardar2007statistical,pathria2011statistical} to analyze possible finite-temperature phase transitions of the 4D X-cube model. The decoupling of the partition function $\mathcal{Z}=\frac{1}{d_h}\mathcal{Z}_{\textsf{hypercube}}\mathcal{Z}_{\textsf{link}}$ means that the free energy can be decoupled as: $F_{\textsf{4D X-cube}}=-\frac{1}{\beta}(\ln\mathcal{Z}_{\textsf{hypercube}}+\ln\mathcal{Z}_{\textsf{link}}-L^4\ln2)=F_{\textsf{hypercube}}+F_{\textsf{link}}+\frac{1}{\beta}L^4\ln2$. Since $F_{\textsf{hypercube}}$ is dual to a 1D Ising model which has no finite-temperature transition, we focus on $F_{\textsf{link}}$ when considering possible finite-temperature phase transition. As we expect the gauge charges of the gauge field $\sigma^x$ are deconfined at zero temperature and confined at high temperature, which can also be diagnosed using 't Hooft loops in the next section, there must be a (zero- or finite-temperature) phase transition in the link part $F_{\textsf{link}}$. Thus, if the absence of zero temperature phase transition can be proved, a finite-temperature phase transition must exist. Therefore, we perform a low-temperature series expansion of the free energy density to see whether non-analytical behavior will emerge in the vicinity of zero temperature.

For excitations in the link part, the operator creating the lowest energy excitation is a $\sigma^z$ on a specific plaquette, which flips 8 link terms and the excitation energy is $16J_B$ \cite{li2020fracton}. 
Since there are $6L^4$ excitations of this kind, we expand $\mathcal{Z}_{\textsf{link}}$  in the vicinity of $T = 0$ as follows \footnote{Concretely, the expansion of the partition function is \cite{li2019finite} 
$\mathcal{Z}_{\textsf{link}}(\beta)=\sum_{\{s_i\}}e^{-\beta H_{\textsf{link}}(\{s_i\})}
     =2^{L^4}\sum_{\{B_l^{\mu}=\pm1\}}e^{-\beta H_{\textsf{link}}(\{B_l^{\mu}\})}
     =2^{L^4} e^{12\beta L^4}(1+6L^4 e^{-16\beta}+\cdots).$
The constant $2^{L^4}$ appears when we change the summation of $\{s_i\}$ to $\{B_l^{\mu}\}$. A specific $\{B_l^{\mu}\}$ configuration corresponds to $2^{L^4}$ $\{s_i\}$ configuration due to the local gauge transformation.}:
\begin{equation}\label{expansion}
     \mathcal{Z}_{\textsf{link}}(\beta)=2^{L^4} e^{-\beta E_0}(1+6L^4 e^{-16\beta}+\cdots),
\end{equation}
where $E_0=-12L^4$ is the ground-state energy of the link Hamiltonian and $\cdots$ represents contributions from higher excited states. 
The free-energy density (per site) $f_\textsf{link}(\beta)=-\frac{1}{\beta L^4}\ln \mathcal{Z}_{\textsf{link}}(\beta)$ is given as:
\begin{eqnarray}\label{freeenergy}
 f_\textsf{link}(\beta) = -\frac{1}{\beta}\left(\ln 2+12\beta+6 e^{-16\beta}+\cdots \right).~~
\end{eqnarray}
$\beta(f_\textsf{link}-f_\textsf{link}(T=0))$ is analytic with respect to $e^{-16\beta}$, up to the dominant terms in the low temperature regime, which indicates there is no zero temperature phase transition in the link sector. 

\textit{The deconfinement-confinement transition.---}We now show that this finite-temperature phase transition discussed above is a deconfinement-confinement transition. This result can be seen from different behaviour of the expectation value of a 't Hooft loop $T_{\gamma}=\prod_{r \in \gamma} \sigma^{x}(r)$ in high/low temperature  \cite{fradkin2013field,RevModPhys.51.659}:

\begin{equation}
    \left\langle T_{\gamma}\right\rangle 
    = \frac{1}{\mathcal{Z}_{\textsf{link}}} \operatorname{Tr}\left[T_{\gamma} \prod_{l\mu}\left(\cosh \beta+\mathcal{B}_{l}^{\mu} \sinh \beta\right)\right],
    \end{equation}
 where $\gamma$ is formed by centers of the plaquettes along any closed contour as shown in Fig.~\ref{loop}.

In the high temperature regime $\beta\ll 1$, the leading contribution of the denominator $\mathcal{Z}_{\textsf{link}}$ is the product of all the $\cosh\beta$ terms. And the leading contribution to the numerator is the production of $\mathcal{B}^{\mu}_l\sinh{\beta}$ terms inside $\Sigma$, where $\Sigma$ is the minimum surface with boundary $\gamma$, and $\cosh\beta$ terms elsewhere. Thus, we obtain  $\left\langle T_{\gamma}\right\rangle$ as:
\begin{equation}\label{dqc1}
\begin{aligned}
    \left\langle T_{\gamma}\right\rangle\approx\tanh{\beta}^{S[\Sigma]}
    =\exp (-\ln(1/\tanh{\beta})S[\Sigma]), 
    \end{aligned}
    \end{equation}
This is exactly the area law of the 't Hooft loop, which indicates that the 
phase at high temperature is confined. 

In the low-temperature regime $\beta\gg 1$, we can assume the creation operators of lowest energy excitation in the Eq. (\ref{expansion}) are well separated and approximately independent, which is known as the dilute limit.  Thus, the low-temperature expansion of partition function is $\mathcal{Z}_\textsf{link}\approx 2^{L^4}e^{-\beta E_0}\sum_{n=0}\text{C}_{N_p}^n e^{-16\beta n}$, where $N_p=6L^4$ is the number of total plaquettes.
In the thermodynamic limit, the binomial coefficient can be approximated to $\frac{N_p^n}{n!}$. Thus the partition function is:
$\mathcal{Z}_{\textsf{link}}\approx 2^{L^4}e^{-\beta E_0}\sum_{n=0} \frac{N_p^n}{n!}e^{-16\beta n} = 2^{L^4}e^{-E_0\beta}e^{N_p e^{-16\beta}}$.
As for the numerator, once a $\sigma^{z}$ plaquette is located on the 't Hooft loop contour $\gamma$, the expectation value of the 't Hooft loop changes its sign. Thus, we obtain $\left\langle T_{\gamma}\right\rangle$ as:
\begin{eqnarray}\label{dqc2}
\begin{aligned}
    \langle T_{\gamma}\rangle &\approx \frac{2^{L^4}\langle T_{\gamma}\rangle_\textsf{gs}}{\mathcal{Z}_{\textsf{link}}}e^{-\beta E_0}\sum_{n=0}\frac{(N_p-2P)^n}{n!}e^{-16\beta n}\\
    &= \langle T_{\gamma}\rangle_\textsf{gs}\frac{e^{-\beta E_0}e^{e^{-16\beta}(N_p-2P)}}{e^{-\beta E_0}e^{e^{-16\beta}N_p}} = e^{-2e^{-16\beta} P},
    \end{aligned}
\end{eqnarray}
where $P$ is the length of the 't Hooft loop and $\langle T_{\gamma}\rangle_\textsf{gs}$ is the ground-state expectation value of $T_{\gamma}$ that equals to one, which is consistent with the physical intuition that local operators cannot distinguish different ground states. The calculation of $\langle T_{\gamma}\rangle_\textsf{gs}$ is shown in the SM.  Thus, the expectation value of the 't Hooft loop operator $\langle T_{\gamma}\rangle$ obeys the perimeter law in the low-temperature regime, which indicates a deconfined phase. Consequently, there must be a deconfinement-confinement phase transition at finite temperature for the 4D X-cube model.

\textit{Spontaneous symmetry breaking of $\mathbb{Z}_2$ n-form subsystem symmetry.---}Here we show that the finite-temperature phase transition of the 4D X-cube model can be physically understood as SSB of $\mathbb{Z}_2$ 1-form subsystem symmetry. For an ordinary $\mathbb{Z}_2$ $n$-form global symmetry, the charged operators are  $n$-dimensional objects \footnote{In this language, usual symmetries such as the $\mathbb{Z}_2$ symmetry in the Ising model are 0-form symmetries} , and are classified by $H^n(M,\mathbb{Z}_2)$ \cite{Gaiotto:2014kfa,Jian:2020eao,lake2018higherform,PhysRevB.99.014402}, where $M$ is the spatial base manifold.
Then a subsystem $n$-form $d_s$-dimensional symmetry (SNS) is defined as the union of all the $n$-form transformations 
in different $d_s$-dimensional subsystems of the original $D$-dimensional lattice  with $n$\textless $d_s$$\textless$$D$ \footnote{In this letter, we consider the lattice model with periodic boundary condition and the associated subsystems are  tori. For example, if a subsystem is $\mathbb{T}^1$, then corresponding elements act only on this 1 dimensional submanifold and this kind of symmetry is also called line symmetry}\cite{PhysRevB.98.035112,PhysRevB.98.235121,PhysRevResearch.2.012059}, and the number of generators depend on the lattice size. 
In this Letter,  the associated subsystems are tori and the subgroup of a $\mathbb{Z}_2$ $n$-form $d_s$-dimensional subsystem symmetry in each torus $\mathbb{T}^{d_s}$   is classified by the $n$-th cohomology group $H^{n}(\mathbb{T}^{d_s},\mathbb{Z}_2)$. 
Physically, a $\mathbb{Z}_2$ $n$-form $d_s$-dimensional subsystem transformation is the same as the ordinary $n$-form transformation in each torus $\mathbb{T}^{d_s}$ subsystem and the generators are $d_s - n$ dimensional objects .

The 4D X-cube model has a 1-form subsystem symmetry and a 2-form subsystem symmetry in each three-dimensional subsystem (namely $d_s=3$).  They are generated by the non-contractible Wilson loops $W$ and 't Hooft loops $T$, respectively, as illustrated in the Fig. \ref{topological}.  
The Wilson loops and 't Hooft loops are 
order parameters of each other when they intersect and anti-commute. Due to the nature of ‘gauge-like symmetry’, the expectation value of the order parameters has the so-called ‘dimensional reduction’ properties, and is summarized as a generalized Elitzur's theorem in \cite{C2005Generalized}. For example, the expectation value $\langle W\rangle$, is bounded from above by the expectation value of $\sigma_z$ on the plaquette, where the $W$ and $T$ intersect, computed in a 
1D system. Here the 1D system consists of the spins acted by $T$ and  in this 1D system $T$ now becomes a  global 0-form $\mathbb{Z}_2$ symmetry, which cannot be SSB at finite temperature. As a result, $\langle W\rangle$ must be zero at any finite temperature, which implies that the $T$ symmetry must be restored at any finite temperature. Similarly, the expectation value of $T$ is bounded from above by that of an order parameter of a 2D system where $W$ symmetry becomes a global 0-form $\mathbb{Z}_2$ symmetry and it is possible to be broken at low but finite temperature. This agrees with the result discussed in the previous sections where only the link terms can support a finite-temperature phase transition. Thus, below $T_c$ the $W$ symmetry is broken and the 4D X-cube model has finite-temperature classical fracton topological order \cite{PhysRevLett.107.210501,castelnovo2007entanglement}. The details of the subsystem and upper bound construction are left in the 'Generalized Elitzur's theorem' section in SM.

Finally, the finite temperature physics of the 4D X-cube model is totally different form the 4D toric code model. The 4D toric code model has two anti-commuting global 2-form symmetries and  both two symmetries are SSB at finite temperature. Since the thermal phase transtion of the 4D X-cube model is due to the SSB of the 1-form subsystem symmetry, its universality class is different from that of the 4D toric code model. Finally, the lower critical dimension of the 2-form subsystem and ordinary global symmetries are different, as we will show in the no-go theorem in the subsequent section. 

\textit{No-go theorem.---}
Similar to the no-go theorem for 2D stabilizer code and 3D STS models \cite{Bravyi_2009,YOSHIDA20112566}, we can also argue a no-go theorem for quantum FTFTO, in 4D gapped fracton models with two anti-commuting discrete SNS whose charges are $n_1$ and $n_2$ dimensional objects. For 4D gapped fracton models, $(n_1,n_2)$ 
can only be $(1,1)$ or $(2,1)$ \footnote{This is because the dimension of a region on which the two subsystem-symmetry generators act can only be (1,1) or (2,1)}; both cases have a subsystem symmetry with $n=1$. Using the ‘dimensional reduction’ approach, it can be shown that this symmetry must be restored at any finite temperature. As a result, 4D quantum fracton topological order is absent at finite temperature under the above assumption. Quantum FTFTO can be realized with two anti-commuting 2-form subsystem symmetries, and this can only be realized in at least 5D. This result can be generalized to models in any spatial dimension as along as the fracton model has a subsystem symmetry whose generators are one dimensional objects, and the charge operators of each subsystem $\mathbb{T}^{d_s}$ belonging to the cohomology group $H^{d_s-1}(\mathbb{T}^{d_s},\mathbb{Z}_{m})$ . As concrete examples, we discuss a series of fracton models in general spatial dimension with subsystem-symmetry generators of each subsystem $\mathbb{T}^3$ :  $(H^1(\mathbb{T}^3,\mathbb{Z}_2),H^2(\mathbb{T}^3,\mathbb{Z}_2))$. 

\textit{Higher-dimensional fracton models.---}The above discussion on the 4D X-cube model can be generalized to a series of higher-dimensional models, summarized as a family tree \cite{li2020fracton,li2021fracton}. The family tree consists of fracton models in spatial $D$ dimensions which are labeled by four indices 
$[d_1,d_2,d_3,D]$.
Here $d_2$ is the dimension of the cube where spins live on. The Hamiltonian contains two terms: one is defined on the $D$ dimensional cube, and the other is defined on the $d_1$ dimensional cube labeled by $\gamma_{d_1}$. Besides, the index $d_3$ means the second term only includes the spins in the $d_3$ dimensional leaf space associated with a given $\gamma_{d_1}$. The partition function of the series $[0,1,2,D]$ is dual to 0d and 1d systems, which has only zero temperature phase transitions. The series $[1,2,3,D]$ has the same SNS as the 4D X-cube model. And the free energy and 't Hooft loop calculation shows there exists a finite temperature phase transition in this series of models. The calculation of partition function, free energy and 't Hooft loops is left in SM.


\textit{Concluding remarks.---}In this Letter, we have shown the existence of finite-temperature classical fracton topological order in 4D X-cube model and, in general, a series of fracton models dubbed as $[1,2,3,D]$ ($D\ge 4$). 
We also argued a no-go theorem for quantum FTFTO given the SNS which only acts on one-dimensional regions. We believe that this provides an important first step towards quantum FTFTO and 
SCQM using fracton models. 


{\it Acknowledgement:} We sincerely thank Xuan Zou on related collaborations. This work was supported in part by NSFC under Grant No. 11825404 (ZZW and HY), the MOSTC under Grants No. 2018YFA0305604 and 2021YFA1400100 (HY), and the Strategic Priority Research Program of Chinese Academy of Sciences under Grant No. XDB28000000 (HY).

\bibliography{bib} 
\newpage
\begin{widetext}
	\section{Supplementary Material}
\subsection{A. Partition function duality of the 4D X-cube model\label{A}}
The partition function of the 4D X-cube model is: 
\begin{equation}
\begin{aligned}
    \mathcal{Z}  &= \operatorname{Tr}\left[\prod_{hc}\left(\mathbb{I} \cosh\beta+\mathcal{A}_{hc} \sinh\beta\right) \prod_{l}^{}\prod_{\mu}\left(\mathbb{I} \cosh\beta +B_{l}^\mu \sinh\beta\right)\right]\\
    & =\frac{1}{d} \operatorname{Tr}\left[\prod_{hc}^{}\left(\mathbb{I} \cosh\beta+\mathcal{A}_{hc} \sinh\beta\right) \right] \operatorname{Tr}\left[ \prod_{l}^{}\prod_{\mu}\left(\mathbb{I} \cosh\beta +B_{l}^\mu \sinh\beta\right)\right]\\
      &=\frac{1}{d}\mathcal{Z}_{\textsf{hypercube}}\times\mathcal{Z}_{\textsf{link}},
    \end{aligned}
     \label{partition}
\end{equation}
     
Next let's prove the duality of of the type-I $\mathcal{Z}_{\textsf{link(type-I)}} $ and type-II $\mathcal{Z}_{\textsf{link(type-II)}} $ contribution to the$\mathcal{Z}_{\textsf{link}}$.  
$\mathcal{Z}_{\textsf{link(type-I)}} $ is given as follows:
 \begin{equation}
\begin{aligned}
   \mathcal{Z}_{\textsf{link(type-I)}}& = \cosh\beta^{12(L-1)^4}\text{Tr}\left[\prod_l\prod_{\mu}\left(\mathbb{I} +\mathcal{B}_{l}^\mu \tanh\beta\right)\right]\\  
   & = d_h\cosh\beta^{12(L-1)^4}\sum_{n=0}^{4(L-1)^{4}}\left(\begin{array}{c}
        4(L-1)^{4} \\
        n
        \end{array}\right)\tanh\beta^{3n}\\
        & = d_h\cosh\beta^{12(L-1)^4}(1+\tanh\beta^3)^{4(L-1)^4}\\
        &= d_h [(\cosh\beta)^3+(\sinh\beta)^3]^{4(L-1)^4}
    \end{aligned}.\\
\end{equation}
Since the term $[(\cosh\beta)^3+(\sinh\beta)^3]^{4(L-1)^4}$ is just the partition function of $4(L-1)^4$ copies of classical 0D system with three Ising interaction: 
$
\hat{H}_{\textsf{3-spin Curie-Weiss}}=-s_1s_2-s_2s_3-s_3s_1 
$, $\mathcal{Z}_{\textsf{link(type-I)}} $
is dual to decoupled 3-spin Curie-Weiss model.

As shown in the Fig. \ref{duality}, type-II constraints are dual to plaquette constraints in the 3D toric code model. We can first consider the partition function of the 3D toric code model which is defined on a $L\times L \times L$ lattice under OBC. The partition function of the 3D toric code model is also decomposed as: $ \mathcal{Z}_{\textsf{3DTC}}=\frac{1}{d_{TC}}\mathcal{Z}_{\textsf{star}}\times\mathcal{Z}_{\textsf{link}},d_{TC}=\text{Tr}[\mathbb{I}]$. The star part $ \mathcal{Z}_{\mathcal{A}_{s}}$ is dual to 1D Ising model $\cosh{\beta}^{(L-1)^3}$. And the plaquette part $ \mathcal{Z}_{\mathcal{B}_{p}}$ is
\begin{equation}
\begin{aligned}
    \mathcal{Z}_{\mathcal{B}_{p}}  =\operatorname{Tr}\left[\prod_p \left(\mathbb{I}\cosh\beta+\mathcal{B}_{p}\sinh\beta\right)\right] 
    =d_{TC}\cosh\beta^{3(L-1)^3}\sum_{\{p\}} \tanh\beta^{n(\{p\})},\\
\end{aligned}
\end{equation}
 where $\{p\}$ is a certain plaquette constraint, and $n(\{p\})$ is the number of plaquette terms of this constraint.
 
 Meanwhile, the $ \mathcal{Z}_{\textsf{link(type-II)}}$ part of the 4D X-cube model is:
\begin{equation}
\begin{aligned}
       \mathcal{Z}_{\textsf{link(type-II)}}  = d_h(\cosh\beta)^{12(L-1)^4} \prod_{i=1}^{4}\sum_{\{p_i\}}(\tanh\beta) ^{n(\{p_i\})},\\
       \end{aligned}
\end{equation}
 where $\{p_i\}$ is a certain type-II constraint, and $n(\{p_i\})$ is the number of link terms of this constraint.
 
 Thus $\mathcal{Z}_{\textsf{hypercube}}\times\mathcal{Z}_{\textsf{link(type-II)}} $ is dual to $4L$ decoupled 3D toric code partition function $\mathcal{Z}_{\mathsf{TC}}$, due to the foliated structure of the 4D X-cube model \cite{10.21468/SciPostPhys.6.4.043,PhysRevLett.126.101603,SHIRLEY2019167922,PhysRevB.102.115103}, divided by a 1D Ising model:
 \begin{equation}
    \mathcal{Z}_{\textsf{hypercube}}\times\mathcal{Z}_{\textsf{link(type-II)}}\sim( \mathcal{Z}_{\mathsf{TC}})^{4L}/(\mathcal{Z}_{\mathsf{1D Ising}}),
\end{equation}
where the length of the 1D Ising model is $3L^4$.

We further prove there is no other local constraints. First, let's only consider the type I constraints under which each link has two independent link terms.  Thus for each vertex, there are four links and eight independent link terms belonging to it. Moreover, we note that type I and II constraints are not independent. Written explicitly, the type-I constraints of a specific vertex are:
\begin{equation}
    \begin{aligned}
   &\mathcal{B}_{x}^y \mathcal{B}_{x}^z \mathcal{B}_{x}^w =\mathcal{B}_{
  \bar{x}}^y \mathcal{B}_{\bar{x}}^z \mathcal{B}_{\bar{x}}^w  =\mathbb{I}\\
    &\mathcal{B}_{y}^x \mathcal{B}_{y}^z \mathcal{B}_{y}^w =\mathcal{B}_{\bar{y}}^x \mathcal{B}_{\bar{y}}^z \mathcal{B}_{\bar{y}}^w=\mathbb{I}\\
    &\mathcal{B}_{z}^x \mathcal{B}_{z}^y \mathcal{B}_{z}^w =\mathcal{B}_{\bar{z}}^x \mathcal{B}_{\bar{z}}^y \mathcal{B}_{\bar{z}}^w= \mathbb{I}\\
    &\mathcal{B}_{w}^x \mathcal{B}_{w}^y \mathcal{B}_{w}^z =\mathcal{B}_{\bar{w}}^x \mathcal{B}_{\bar{w}}^y \mathcal{B}_{\bar{w}}^z= \mathbb{I}.\\
    \end{aligned}
\end{equation}
There are only four independent type-I constraints in the above eight type-I constraints, as each link has two vertices. The type-II constraints are:
\begin{equation}
\begin{aligned}
   &\mathcal{B}_{x}^w \mathcal{B}_{y}^w \mathcal{B}_{z}^w \mathcal{B}_{\bar{x}}^w \mathcal{B}_{\bar{y}}^w \mathcal{B}_{\bar{z}}^w = \mathbb{I}\\
    &\mathcal{B}_{x}^z \mathcal{B}_{y}^z \mathcal{B}_{w}^z \mathcal{B}_{\bar{x}}^z \mathcal{B}_{\bar{y}}^z \mathcal{B}_{\bar{w}}^z = \mathbb{I}\\
   &\mathcal{B}_{x}^y \mathcal{B}_{z}^y \mathcal{B}_{w}^y \mathcal{B}_{\bar{x}}^y \mathcal{B}_{\bar{z}}^y \mathcal{B}_{\bar{w}}^y = \mathbb{I}\\
    &\mathcal{B}_{y}^x \mathcal{B}_{z}^x \mathcal{B}_{w}^x \mathcal{B}_{\bar{y}}^x \mathcal{B}_{\bar{z}}^x \mathcal{B}_{\bar{w}}^x = \mathbb{I}\label{last}.
\end{aligned}
\end{equation}
We can obtain the last type II constraint from the type I constraints and the first three type II  constraints. Therefore, there are three independent type II constraints for each vertex, which implies five independent link terms for each vertex. What's more, all the $L^4$ hypercube terms are independent. As a result, there are $6$ independent $\mathbb{Z}_2$ variables per vertex, which is exactly the same as the number of plaquettes per vertex, or the total number of spins.

\subsection{B. The partition function under PBC}
In this section, we prove the partition function under periodic boundary condition (PBC) gives the same phase transition as that under OBC. This is because the free energy density are the same under PBC and OBC in the thermodynamic limit \cite{weinstein2020absence}. We focus on the link sector in this section, and the derivation can be directly applied to the hypercube sector. We use one symbol $l$ to label a link term for simplicity. There are some local constraints $\prod_{j\in s_{k}} \mathcal{B}_j=1$ and $s_{k}$ denotes the set of $\mathcal{B}_l$ belonging to the local constriants. Thus the partition function under OBC is
\begin{equation}
\begin{aligned}
   \mathcal{Z}_{\text{OBC}} 
 = \sum_{\text{OBC},\{s_k\}}\cosh^{N}\beta(\tanh\beta)^{n({\{s_k\}})},
   \end{aligned}
\end{equation}
where the sum is over all the possible constraint  configurations $\{s_k\}$, and $n({\{s_k\}})$ is the number of $\mathcal{B}_l$ terms under the configuration $\{s_k\}$.

We first assume only one global constraint $\prod_{j\in S} \mathcal{B}_j=1$  under PBC, and the number of $\mathcal{B}_l$ terms  in the set $S$ $n_{S}$ is O($L$) at least. 
Thus the partition function with this constraint is \begin{equation}
\begin{aligned}
   \mathcal{Z} & =\operatorname{Tr}_{\text{OBC}}\left(\frac{1+\prod_{j\in S} \mathcal{B}_j}{2}e^{-\beta H_B}\right)\\
   & =\frac{1}{2}\mathcal{Z}_{\text{OBC}}+\frac{1}{2}\cosh^{N}\beta\tanh^{n_S}\beta\sum_{\text{OBC},\{s_k\},s_k \cap S=\emptyset,}\tanh^{n(\{s_k\})}\beta.
   \end{aligned}
\end{equation}
In the thermodynamic limit, the ratio between the second term and the first term is zero. Thus free energy is $F=F_{\text{OBC}}+\frac{\ln 2}{\beta}$.

Now, we assume there are $M$ independent global constraints whose orders of magnitude is O($L^{3}$) at most. This is due to the fact they are all boundary effects. Then the density of free energy for PBC is \begin{equation}
\begin{aligned}
   f_{\text{PBC}}=f_{\text{OBC}}+M\frac{\ln 2}{6L^{4}\beta}.
   \end{aligned}
\end{equation}
Thus the density of free energy in the PBC is the same as that of OBC in the thermodynamic limit.

\subsection{C. Field theory description}

\subsubsection{1. Effective field theory derivation}
In this section, we will give an effective field theory description of the 4D X-cube model. We follow the procedure of the 3D X-cube model \cite{slagle2017quantum}, and start from the 4D $\mathbb{Z}_N$ X-cube model, which is a natural generalization of the original $\mathbb{Z}_2$ X-cube model. The model is defined by the following Hamiltonian:
\begin{equation}
  \mathcal{H}_{\mathrm{X}-\mathrm{cube}}=-\sum_{hc}\left(\hat{\mathcal{A}}_{hc}+\hat{\mathcal{A}}_{hc}^{\dagger}\right)-\sum_{l, a}\left(\hat{\mathcal{B}}_{l}^{a}+\hat{\mathcal{B}}_{l}^{a\dagger}\right),
    \end{equation}
      where $\hat{\mathcal{A}}_{hc}$ and $\hat{\mathcal{B}}_{l}^{a}$ are defined as tensor product of the $\mathbb{Z}_N$ version of Pauli matrices $\hat{Z}_{i}$ and $\hat{X}_{i}$ :  
\begin{align}\label{commutation}
\hat{X}_{i} \hat{Z}_{j}  =\omega^{\delta_{i j}} \hat{Z}_{j}\hat{X}_{i}, \ 
\omega  =e^{2 \pi i / N} .
\end{align}
For examples, the $\hat{\mathcal{A}}_{hc}$ and $\hat{\mathcal{B}}_{l}^{a}$ term 
for the special hypercube (1/2,1/2,1/2,1/2) and link (0,0,0,1/2) are given as follows:
\begin{align}\label{Z_{N} Hamiltonian}
\hat{\mathcal{A}}_{(\frac{1}{2},\frac{1}{2},\frac{1}{2},\frac{1}{2})}=&\hat{Z}_{(0,0,\frac{1}{2},\frac{1}{2})} \hat{Z}^{\dagger}_{(1,0,\frac{1}{2},\frac{1}{2})}\hat{Z}^{\dagger}_{(0,1,\frac{1}{2},\frac{1}{2})}\hat{Z}_{(1,1,\frac{1}{2},\frac{1}{2})}\nonumber\\&\hat{Z}_{(0,\frac{1}{2},0,\frac{1}{2})}  \hat{Z}^{\dagger}_{(1,\frac{1}{2},0,\frac{1}{2})}\hat{Z}^{\dagger}_{(0,\frac{1}{2},1,\frac{1}{2})}\hat{Z}_{(1,\frac{1}{2},1,\frac{1}{2})}\nonumber\\&\hat{Z}_{(0,\frac{1}{2},\frac{1}{2},0)}  \hat{Z}^{\dagger}_{(1,\frac{1}{2},\frac{1}{2},0)}\hat{Z}^{\dagger}_{(0,\frac{1}{2},\frac{1}{2},1)}\hat{Z}_{(1,\frac{1}{2},\frac{1}{2},1)}\nonumber\\
&\hat{Z}_{(\frac{1}{2},0,0,\frac{1}{2})} \hat{Z}^{\dagger}_{(\frac{1}{2},0,1,\frac{1}{2})}\hat{Z}^{\dagger}_{(\frac{1}{2},1,0,\frac{1}{2})}\hat{Z}_{(\frac{1}{2},1,1,\frac{1}{2})}\nonumber\\&\hat{Z}_{(\frac{1}{2},0,\frac{1}{2},0)}  \hat{Z}^{\dagger}_{(\frac{1}{2},0,\frac{1}{2},1)}\hat{Z}^{\dagger}_{(\frac{1}{2},1,\frac{1}{2},0)}\hat{Z}_{(\frac{1}{2},1,\frac{1}{2},1)}\nonumber\\&\hat{Z}_{(\frac{1}{2},\frac{1}{2},0,0)}  \hat{Z}^{\dagger}_{(\frac{1}{2},\frac{1}{2},1,0)}\hat{Z}^{\dagger}_{(\frac{1}{2},\frac{1}{2},0,1)}\hat{Z}_{(\frac{1}{2},\frac{1}{2},1,1)},\\
\hat{\mathcal{B}}^{3}_{(0,0,0,\frac{1}{2})}=&\hat{X}_{(-\frac{1}{2},0,0,\frac{1}{2})}\hat{X}^{\dagger}_{(\frac{1}{2},0,0,\frac{1}{2})}\hat{X}^{\dagger}_{(0,-\frac{1}{2},0,\frac{1}{2})}\hat{X}_{(0,\frac{1}{2},0,\frac{1}{2})},\\
\hat{\mathcal{B}}^{2}_{(0,0,0,\frac{1}{2})}=&\hat{X}^{\dagger}_{(-\frac{1}{2},0,0,\frac{1}{2})}\hat{X}_{(\frac{1}{2},0,0,\frac{1}{2})}\hat{X}_{(0,0,-\frac{1}{2},\frac{1}{2})}\hat{X}^{\dagger}_{(0,0,\frac{1}{2},\frac{1}{2})},\\
\hat{\mathcal{B}}^{1}_{(0,0,0,\frac{1}{2})}=&\hat{X}_{(0,-\frac{1}{2},0,\frac{1}{2})}\hat{X}^{\dagger}_{(0,\frac{1}{2},0,\frac{1}{2})}\hat{X}^{\dagger}_{(0,0,-\frac{1}{2},\frac{1}{2})}\hat{X}_{(0,0,\frac{1}{2},\frac{1}{2})}.
\end{align}
The mapping from the lattice operators to the field operators in the effective field theory is:

\begin{equation}\label{densities}
\begin{aligned} 
\hat{Z}_{i}(t) & \sim \exp \left(i \int_{S}  Z_{ab}(\boldsymbol{x}, t))\right),
\hat{X}_{i}(t) \sim \exp \left(i \int_{\perp S} X_{ab}(\boldsymbol{x}, t)\right), \\ \hat{\mathcal{A}}_{hc}(t) & \sim \exp \left(\frac{2 \pi i}{N} \int_{hc} i^{0}(\boldsymbol{x}, t)\right) , \hat{\mathcal{B}}_{a}^{b}(\boldsymbol{r}, t)  \sim \exp \left(\frac{2 \pi i}{N} \int_c j^{0 ; ab}(\boldsymbol{x}, t)\right) ,
\end{aligned}
\end{equation}
    where the integration regime $S$ is the plaquette where operator $\hat{Z}_{i}$ lives on and $\perp S$ is the dual plaquette of $S$. The regime $hc$ is the hypercube where $\hat{\mathcal{A}}_i(t) $ is centered in and $c$ is the cube which is the dual space of the link $a$. 
    
    Next, we show how to interpret the spatial indices $a,b$ of the gauge fields. $a,b$ are the bases vectors direction of the plaquette where the gauge field lives. Thus the gauge fields are symmetric rank-2 tensor gauge fields and we denote them as $Z_{\rho\sigma}(x,t)$ and $X_{\rho\sigma}(x,t)$ with $\rho,\sigma\in \{1,2,3,4\}$, where we use $ \{1,2,3,4\}$ to represent the spatial directions $\{x,y,z,w\}$ for simplicity. For the sake of brevity, let's assume that $\rho \textless \sigma$ in any summation of the indices, but we retain the general $\rho$ and $\sigma$ otherwise to illustrate the symmetric properties of the gauge fields explicitly.\par 
   Since $\hat{Z}$ and $\hat{X}$ have the commutation relation (\ref{commutation}), the field $Z_{\rho\sigma}(x,t)$ and $X_{\rho\sigma}(x,t)$ should have the equal-time commutation relation as follows:
     \begin{equation}
        \left[Z_{\alpha\beta}(t, \boldsymbol{x}), X_{\gamma\theta}\left(t, \boldsymbol{x}^{\prime}\right)\right]=\frac{2 \pi i}{N} (\delta_{\alpha\gamma}\delta_{\beta\theta}+\delta_{\alpha\theta}\delta_{\beta\gamma}) \delta^{4}\left(\boldsymbol{x}-\boldsymbol{x}^{\prime}\right).
        \end{equation}
    
   And $i^0$ and $j^0$ are fracton and dimension-2 strings densities which can be directly read off from the Hamiltonian:
    \begin{eqnarray}
   &&i^{0}=\frac{N}{2 \pi}\left|\epsilon^{0 \alpha \beta \gamma \delta}\right| \partial_{\alpha} \partial_{\beta} Z_{\gamma \delta},\nonumber\\
   &&j^{0 ; \rho \sigma}=\frac{N}{2 \pi} \epsilon^{0 \rho \sigma \mu \nu} \partial_{\nu} X^{\nu \rho},
    \end{eqnarray}
    where $\epsilon^{0\alpha\beta \gamma\delta}$ is the 5-order Levi-Civita symbol, and the absolute value sign here is to make sure the indices are different. 
    The densities of the dimension-2 string $j^{0 ; \rho \sigma}$ has 12 components. The first spatial label $\rho$ is the direction of the link $l$ which takes values freely from 1 to 4, and the second component $\sigma$ take the three values except $l$. 
 
Using the expression of densities, we can construct the Lagrangian of the 4D $\mathbb{Z}_N$ X-cube model. Similar to the construction of the BF field theory, the Lagrangian density is:
\begin{equation}
    \begin{aligned}
    \mathcal{L}_{\mathrm{X}-\mathrm{cube}} &=\frac{N}{2 \pi} X^{\alpha \beta} \partial_{0} Z_{\alpha\beta} +X_{0} 
    \underbrace{\frac{N}{2 \pi}\left|\epsilon^{0\alpha\beta\gamma\delta}\right|\partial_{\alpha} \partial_{\beta} Z_{\gamma\delta}}_{i^{0}}\\
    &+Z_{0 ; \alpha\beta} \underbrace{\frac{N}{2 \pi} \epsilon^{0 \alpha\beta\gamma\delta} \partial_{\gamma} X^{\gamma\alpha}}_{j^{0 ; \alpha \beta}} 
    -Z_{0 ; \alpha\beta} j^{0 ; \alpha\beta}-|\epsilon^{\alpha\beta\gamma\delta}|Z_{\gamma\delta} j^{\alpha\beta}-X_{0} i^{0}-|\epsilon^{\alpha\beta\gamma\delta}|X_{\gamma\delta} i^{\alpha\beta}.
    \end{aligned}
    \end{equation} 
We redefine the fields as:
\begin{equation}
    X_{\alpha\beta} = B_{\gamma\delta} |\epsilon^{0\alpha\beta\gamma\delta}|\quad, Z_{\alpha\beta} = A_{\gamma \delta}|\epsilon^{0\alpha\beta\gamma\delta}|,X_{0}=B_0,Z_{0,\alpha\beta}=A_{0,\alpha,\beta}.
\end{equation}
Thus the Lagrangian density is rewritten as:
\begin{eqnarray}\label{BF}
    \mathcal{L}_{\mathrm{X}-\mathrm{cube}} &&= \frac{N}{2\pi}A_{\alpha\beta}\partial_{0}B_{\alpha\beta}+B_{0}(\frac{N}{2\pi}\partial_{\alpha}\partial_{\beta}A_{\alpha\beta}-i^0)\nonumber\\
    &&+A_{0;\alpha\beta}(\frac{N}{2\pi}\epsilon^{0\alpha\beta\gamma\delta}\partial_{\gamma}B_{\beta\delta}-j^{0;\alpha\beta})-A_{\alpha\beta}j^{\alpha\beta}-B_{\alpha\beta} i^{\alpha\beta}.
    \end{eqnarray}
Here the gauge fields $A$ and $B$ are still the rank-2 tensor gauge fields. The time components are denoted as $A_{0;\alpha\beta}$ and $B_0$, and the spatial components are denoted as $A_{\alpha\beta}$ and $B_{\alpha\beta}$.

\subsubsection{2. Non-contractible Wilson loops, ’t Hooft loops and ground state degeneracy }
In this section, we construct gauge invariant non-contractible Wilson surfaces and 't Hooft loops from the field theory derived above which can give the ground state degeneracy of this model. These field theory constructions in lower dimensional fracton models have been discussed in \cite{10.21468/SciPostPhys.10.2.027,Seiberg:2020cxy,Seiberg:2020wsg}.
From the Eq.(\ref{densities}), we know the density describing the ground states should satisfy the following equations:
\begin{eqnarray}
&&i^{0}=\frac{N}{2\pi}\partial_{\alpha}\partial_{\beta}A_{\alpha\beta}=0,\label{I}\\
&&j^{0,\alpha\beta}=\frac{N}{2\pi}\epsilon^{0\alpha\beta\gamma\delta}\partial_{\gamma}B_{\beta\delta}=0.\label{J}
\end{eqnarray}
And the gauge transformation is given as follows:
\begin{eqnarray}
B_{\alpha\beta}\to&& B_{\alpha\beta}+\int_{x'} [B_{\alpha\beta},i_{0}]\chi,\nonumber\\
=&&B_{\alpha\beta}+\partial_{\alpha}\partial_{\beta}\chi\\
A_{\alpha\beta}\to&& A_{\alpha\beta}+\int_{x'} [A_{\alpha\beta},j_{0,\gamma\delta}]f_{\gamma\delta},\nonumber\\
=&&A_{\alpha\beta}+\epsilon^{\alpha\beta\gamma\delta}\partial_{\gamma}f_{\delta\alpha}+\epsilon^{\beta\alpha\gamma\delta}\partial_{\gamma}f_{\delta\beta}.
\end{eqnarray}
To give the correct ground state degeneracy, we should construct all the independent 't Hooft loops and Wilson surfaces which can label ground states. For simplicity, we first consider the indices of $A$ and $B$ are $xy$. The time direction is denoted as $t$.
For the $B$ gauge field, we construct such non-contractible gauge invariant ’t Hooft loops:
\begin{eqnarray}
T^{y}(x_{0})=\exp\left(i\int_{0}^{L_{y}a}\int_{x_{0}}^{x_{0}+a} B_{xy}dxdy\right),\quad x_{0}=1,\cdots,L_{x};\nonumber\\~\\
T^{x}(y_{0})=\exp\left(i\int_{0}^{L_{x}a}\int_{y_{0}}^{y_{0}+a} B_{xy}dxdy\right),\quad y_{0}=1,\cdots,L_{y},\nonumber\\~
 \end{eqnarray}
where the labels $x$ and $y$ are the spatial directions of the correspondence 't Hooft loop. From the equation (\ref{J}), we know
\begin{eqnarray}
&&j_{0,zx}=\partial_{w}B_{xy}-\partial_{y}B_{xw}=0,\nonumber\\\Rightarrow
&&\partial_{w}T^{y}\sim\int_{0}^{L_{y}a}\int_{x_{0}}^{x_{0}+a} \partial_{y}B_{xw}dxdy=0.
\end{eqnarray}
The independence of the $z$ direction can be similarly proved. Thus $T^{y}$ only depends on $x$ and $y$. We can also prove $T^{x}$ only depends on $x$ and $y$ similarly. There are only $L_{x}+L_{y}-1$ independent 't Hooft loops of $B$ since the product of all the $T^{x}$ equals to that of $T^{y}$.

For the gauge field $A$, we construct the following Wilson surfaces:
\begin{eqnarray}
W^{zw}(x_i,y_i)=\exp\left(i\int^{L_z a}_{z=0}\int^{L_w a}_{w=0} A_{xy}(x_i,y_i,z,w)dzdw\right),\end{eqnarray}
where $ x_i=a,\cdots,L_{x}a\ \text{and}\ y_i=a,\cdots,L_{y}a$, and the label $zw$ is the bases vectors direction of the Wilson surfaces.

 However from the equation (\ref{I}), we obtain $\partial_{x}\partial_{y}(\ln W^{zw}(x,y))=0$. Thus there are only $L_{x}+L_{y}-1$ independent $W$ in the $xy$ plane:
 \begin{eqnarray}
\ln W^{zw}(x,y)=f(x)+g(y)=\ln W^{zw}(x,0)+\ln W^{zw}(0,y)-\ln W^{zw}(0,0).
\end{eqnarray}
The commutation relations between $W$ and $T$ are the same as $L_{x}+L_{y}-1$ copies of $\mathbb{Z}_{N}$ Heisenberg algebra:
 \begin{eqnarray}
T^{y}(x)W^{zw}(x,0)=e^{-\frac{2\pi i}{N}}W^{zw}(x,0)T^{y}(x),\nonumber\\ T^{x}(y)W^{zw}(0,y)=e^{-\frac{2\pi i}{N}}W^{zw}(0,y)T^{x}(y).
\end{eqnarray}
Therefore there are $N^{L_{x}+L_{y}-1}$ ground states labeled by the above operators. The calculation for other directions is similar and the total ground state degeneracy is $N^{3L_{x}+3L_{y}+3L_{z}+3L_{w}-6}$. This is consistent with the result calculating from the lattice model directly \cite{li2020fracton,li2021fracton}.
\\
\subsubsection{3. Excitations and immobility}
In this section, we will discuss the excitation of the 4D X-cube model from field theory. We couple gauge fields to currents and charges, and then the Lagrangian density is:
\begin{eqnarray}\label{lag 2}
\mathcal{L}&&= \frac{N}{2\pi}A_{\alpha\beta}\partial_{0}B_{\alpha\beta}+B_{0}(\frac{N}{2\pi}\partial_{\alpha}\partial_{\beta}A_{\alpha\beta}-i^0)\nonumber\\
&&+A_{0;\alpha\beta}(\frac{N}{2\pi}\epsilon^{0\alpha\beta\gamma\delta}\partial_{\gamma}B_{\beta\delta}-j^{0;\alpha\beta})-A_{\alpha\beta}j^{\alpha\beta}-B_{\alpha\beta} i^{\alpha\beta}.
\end{eqnarray}
After integrating out $A_{\alpha\beta}$ and $B_{\alpha\beta}$, we get the equation of motion(EOM) of currents:
\begin{eqnarray}
&&j^{\alpha\beta}=\frac{N}{2\pi}(\partial_{0}B_{\alpha\beta}+\partial_{\alpha}\partial_{\beta}B_{0}),\\
&&i^{\alpha\beta}=-\frac{N}{2\pi}\partial_{0}A_{\alpha\beta}-\frac{N}{2\pi}\epsilon^{0\alpha\beta\gamma\delta}\partial_{\gamma}(A_{0;\beta\delta}-A_{0;\alpha\delta}).
\end{eqnarray}
The conservation law now can be obtain from Eq. (\ref{I}), (\ref{J})
\begin{eqnarray}\label{conservation law}
&&\partial_{0}j^{0,\alpha\beta}=\epsilon^{0\alpha\beta\gamma\delta}\partial_{\gamma}j^{\beta\delta},\\
&&\partial_{0}i^{0}+\partial_{\alpha\beta}i^{\alpha\beta}=0.
\end{eqnarray}
The density describing the fracton excitation at $\vec{r}=0$ is
\begin{eqnarray}
i_{0}=\delta^{4}(x),\ i_{\alpha\beta}=0.
\end{eqnarray}
A solution to this equation is
\begin{eqnarray}
A_{xy}=\frac{2\pi}{N}\theta(x)\theta(y)\delta(z)\delta(w).
\end{eqnarray}
Now we consider the density of the  dimension-2 string excitation on the $xw$-plane is
\begin{eqnarray}
&&j^{0,xz}=-j^{0,xy}=\delta(y)\delta(z)(\delta(w)-\delta(a-w))\theta(x)\theta(a-x),\\ 
&&-j^{0,wz}=j^{0,wy}=\delta(y)\delta(z)(\delta(x)-\delta(a-x))\theta(w)\theta(a-w),\\
&&B_{yz}=\frac{2\pi}{N}\theta(x)\theta(a-x)\delta(z)\delta(y)\theta(w)\theta(a-w),
\end{eqnarray}

Other kinds of excitation are just the combination of several elementary excitations derived above. As a result, when $N=2$, the lowest energy dimension-2 string excitaion is created by an $\sigma_z$ in a specific plaquette. And there are eight link terms flipped by this operator.

Moreover, similar to the 3D X cube model, the fracton excitation of 4D X cube also has the local conservation of dipole moment and thus an isolated charge is incapable of moving. This result can obtain from conservation law \eqref{conservation law}:
\begin{eqnarray}
\frac{d}{dt}P^{\gamma}=\int d^4 x(\partial_{0} i^{0})x^{\gamma} =\int d^4 x x^{\gamma} \partial_{\alpha\beta}i^{\alpha\beta}=\int dn_{\beta}(x^{\gamma}\partial_{\alpha}i^{\alpha\beta}-i^{\gamma\beta})=0.
\end{eqnarray}
This consists with the lattice understanding\cite{li2020fracton,li2021fracton} of the restricted mobility of excitations.

\subsubsection{4. Generalized Elitzur's theorem}
    The higher-form subsystem symmetries here are gauge-like symmetries, which do not act on the whole system. Gauge-like symmetries have the 'dimension reduction' property, and is summarized as a generalized Elitzur's theorem in \cite{C2005Generalized}. In this section, we use the dimensional reduction method \cite{C2005Generalized} to give the lower critical dimension with respect to the higher symmetry in the 4D $\mathbb{Z}_2$ X-cube model.

    First, let's consider the  $W$ symmetry. We choose one element $W^{zw}(x_{0},0)$ and denote the integration regime in it  as $S$, which is the $zw$ plane. And all the $\sigma ^z$ diagonalized bases in $S$ form a set ${\eta}$, and its complementary set is ${\psi}$. The charged operator is the 't Hooft loop $T^{y}(x_{0})$: $W^{zw}(x_0,0)T^{y}(x_{0})(W^{zw}(x_0,0))^{-1}=-T^{y}(x_{0})$. This charged object is an one-dimensional operator. 
    
The ensemble average of the charged operator is:
\begin{equation}
\begin{aligned}
\langle T^{y}(x_{0})\rangle_{h\rightarrow0^{+},L\rightarrow+\infty}&=\lim_{h\rightarrow0^{+},
L\rightarrow+\infty}\frac{\sum_{\psi,\eta} e^{\beta H(\psi,\eta)+h\sum_{\phi_i\in\{\psi\}\cup\{\eta\}}\phi_i}T^{y}_{\phi_i\in\{\eta\}}T^{y}_{\phi_i\in\{\psi\}}}{\sum_{\psi,\eta} e^{\beta H(\psi,\eta)+\beta h\sum_{\phi_i\in\{\psi\}\cup\{\eta\}}\phi_i}}\\
&=\lim_{h\rightarrow0^{+},
L\rightarrow+\infty}\frac{\sum_{\psi} Z_{\psi}e^{\beta h\sum_{\phi_i\in\{\psi\}}\phi_i}T^{y}_{\psi}\sum_{\eta}\frac{T^{y}_{\eta}e^{\beta H(\psi,\eta)+\beta h\sum_{\phi_i\in\{\eta\}}\phi_i}}{Z_{\psi}}}{\sum_{\psi} Z_{\psi}e^{h\sum_{\phi_i\in\{\psi\}}\phi_i}},
\end{aligned}
\end{equation}

where $Z_{\psi}=\sum_{\eta}e^{\beta H(\psi,\eta)+h\sum_{\phi_i\in\{\psi\}\cup\{\eta\}}\phi_i}$. The $\psi,\eta$ denote the $\sigma_z$ diagolized basis configuration,  $T^y_{\psi},T^y_{\eta}$ means the part of ’t Hooft loop in the $\psi$ and $\eta$ regime respectively, and $H(\psi,\eta)$ is the expectation value of under this basis configuration.

Let $\bar{\psi}$ denotes the configuration which maximize $\left|T^{y}_{\psi}\sum_{\eta}\frac{T^{y}_{\eta}e^{\beta H(\psi,\eta)+\beta h\sum_{\phi_i\in\{\eta\}}\phi_i}}{Z_{\psi}}\right|$ . Then we have the upper bound on $\left|\langle T^{y}(x_{0})\rangle_{h\rightarrow0^{+},L\rightarrow+\infty}\right|$:

\begin{equation}
\begin{aligned}
\left|\langle T^{y}(x_{0})\rangle_{h\rightarrow0^{+},L\rightarrow+\infty}\right|&\leq\lim_{h\rightarrow0^{+},L\rightarrow+\infty}\left|T^{y}_{\bar{\psi}}\sum_{\eta}\frac{T^{y}_{\eta}e^{\beta H(\bar{\psi},\eta)+\beta h\sum_{\phi_i\in\{\eta\}}\phi_i}|}{Z_{\psi}}\right|\\
&\leq\lim_{h\rightarrow0^{+},L\rightarrow+\infty}\left|\sum_{\eta}\frac{T^{y}_{\eta}e^{\beta H(\bar{\psi},\eta)+\beta h\sum_{\phi_i\in\{\eta\}}\phi_i}|}{Z_{\psi}}\right|.
\end{aligned}
\end{equation}

The last line is just the expectation value of order parameter of 2D quantum system $H(\bar{\psi},\eta)$ under external field $\bar{\psi}$  at finite temperature. Now the $W$ subsystem symmetry of the 4D X cube model becomes the $\mathbb{Z}_{2}$ 0-form global symmetry in the 2D system $H(\bar{\psi},\eta)$. From the famous Mermin-Wagner theorem, discrete 0-form global symmetry can be spontaneously broken at finite temperature in 2D system which means its order parameter can be nonzero. Thus it's also possible to spontaneously break the $W$ subsystem symmetry in the 4D X cube model.

However, it can be shown that the $T$ subsystem symmetry is unable to be spontaneously broken at finite temperature. This is because the regime acted by $T$ is a line on the lattice, the final reduced system is a 1D system with local interactions. And the subsystem symmetry $T$ becomes the 0-form global $\mathbb{Z}_{2}$ symmetry of 1D system.   According to the Mermin-Wagner theorem, any discrete symmetry cannot be spontaneously broken at finite temperature in 1D system. Therefore the expectation value of the order parameter of symmetry $T$ must become zero as the external field $h$ goes to zero.

However, the 3D X-cube model has two 1-form subsystem symmetries generalized by the Wilson loops and 't Hooft loops which belong to the cohomology group $H^1(\mathbb{T}^2,\mathbb{Z}_2)$. Since both of them act over one dimensional regime,  they cannot be spontaneously broken at finite temperature, which implies the absence of finite temperature phase transition.

\subsubsection{5. Contractible 't Hooft loops}
In this section, we construct contractible 't Hooft loops from the BF field theory and then map them to the lattice. We can also prove their expectation value at zero temperature all equal to one. This is within expectation as the fracton topological order cannot be detected by local operators. 

We can take a special 't Hooft loop on the $yz$ plane as an example and denote it as $T_{\gamma_{yz}}$. In the infrared limit, it is defined as:
\begin{equation}
\begin{aligned}
   T_{\gamma_{yz}}=\exp[i(&\int_{0}^{y_0}\int_{x_{0}}^{x_{0}+a} B_{xy}(x,y,0,w_0)dxdy+\int_{0}^{z_0}\int_{x_{0}}^{x_{0}+a} B_{xz}(x,y_0,z,w_0)dxdz\\&+\int_{y_0}^{0}\int_{x_{0}}^{x_{0}+a}  B_{xy}(x,y,z_0,w_0)dxdy+\int_{z_0}^{0}\int_{x_{0}}^{x_{0}+a} B_{xz}(x,0,z,w_0)dxdz)],
\end{aligned}
\end{equation}
where the lengths of sides of the ’t Hooft loop are  $|y_0|$ and $|z_0|$. We can directly show that this operator is gauge invariant under the gauge transformation $B_{\alpha\beta}\rightarrow B_{\alpha\beta}+\partial_{\alpha}\partial_{\beta}\chi$:
\begin{equation}
\begin{aligned}
    \int_{0}^{y_0}\int_{x_{0}}^{x_{0}+a} B_{xy}(x,y,0,w_0)dxdy\rightarrow&\int_{0}^{y_0}\int_{x_{0}}^{x_{0}+a} B_{xy}(x,y,0,w_0)dxdy+\chi(x_{0}+a,y_0a,0,w_0)-\chi(x_{0}+a,0,0,w_0)\\
    &-\chi(x_{0},y_0a,0,w_0)+\chi(x_{0},0,0,w_0),\\
    \end{aligned}
    \end{equation}
    \begin{equation}
    \begin{aligned}
    \int_{0}^{z_0}\int_{x_{0}}^{x_{0}+a} B_{xz}(x,y_0,z,w_0)dxdz\rightarrow&\int_{0}^{z_0}\int_{x_{0}}^{x_{0}+a} B_{xz}(x,y_0,z,w_0)dxdz+\chi(x_{0}+a,y_0,z_0,w_0)-\chi(x_{0},y_0,z_0,w_0)\\
    &-\chi(x_{0}+a,y_0,0,w_0)+\chi(x_{0},y_0,0,w_0),\\
    \end{aligned}
    \end{equation}
    \begin{equation}
    \begin{aligned}
    \int_{y_0}^{0}\int_{x_{0}}^{x_{0}+a} B_{xy}(x,y,z_0,w_0)dxdy\rightarrow&\int_{y_0}^{0}\int_{x_{0}}^{x_{0}+a} B_{xy}(x,y,z_0,w_0)dxdy+\chi(x_{0}+a,0,z_0,w_0)-\chi(x_{0},0,z_0,w_0)\\
    &-\chi(x_{0}+a,y_0,z_0,u_0)+\chi(x_{0},y_0,z_0,w_0),\\
    \end{aligned}
    \end{equation}
    \begin{equation}
     \begin{aligned}
    \int_{z_0}^{0}\int_{x_{0}}^{x_{0}+a} B_{xz}(x,0,z,w_0)dxdz\rightarrow&\int_{z_0}^{0}\int_{x_{0}}^{x_{0}+a} B_{xz}(x,0,z,w_0)dxdz+\chi(x_{0}+a,0,0,w_0)-\chi(x_{0},0,0,w_0)\\
    &-\chi(x_{0}+a,0,z_0,w_0)+\chi(x_{0},0,z_0,w_0).\\
    \end{aligned}
\end{equation}
Summing up the above four terms, we find the above ’t Hooft loop is gauge invariant in the field theory.
To map the contractible 't Hooft loops onto the lattice, we can use the operator mapping (\ref{densities}). The lattice contractible 't Hooft loop is shown in Figure (\ref{loop}). The gauge transformation on the lattice is $\hat{X}_{\mathbf{x}} \rightarrow \hat{\mathcal{A}}_{\mathbf{x}^{\prime}}^{\dagger} \hat{X}_{\mathbf{x}} \hat{\mathcal{A}}_{\mathbf{x}^{\prime}}$, where $\hat{\mathcal{A}}_{\mathbf{x}^{\prime}}$ is the hypercube term, and $\mathbf{x}^{\prime}$ is the position where $f_{\alpha\beta}\neq0$. If $\gamma$ is a general closed contour on the lattice, then it is formed by multiplication of elementary loops on a plane. The gauge invariance of a contractible loop on any plane can be proved similarly as $T_{\gamma_{yz}}$. Thus, for a general $\gamma$, the contractible 't Hooft loop is gauge invariant.

Next, we can prove that the expectation value of $T_{\gamma_{yz}}$  is one in the ground state:
\begin{equation}
    \begin{aligned}
   &\int_{0}^{y_0}\int_{x_{0}}^{x_{0}+a} B_{xy}(x,y,0,w_0)dxdy+\int_{0}^{z_0}\int_{x_{0}}^{x_{0}+a} B_{xz}(x,y_0,z,w_0)dxdz\\
    &+\int_{y_0}^{0}\int_{x_{0}}^{x_{0}+a} B_{xy}(x,y,z_0,w_0)dxdy+\int_{z_0}^{0}\int_{x_{0}}^{x_{0}+a} B_{xz}(x,0,z,w_0)dxdz)\\
    &=\int_{x_{0}}^{x_{0}+a}dx\int_{\textbf{loop}} dydz(\partial_y B_{xz}-\partial_zB_{xy})\\
    &=0.
\end{aligned}
\end{equation}
The second equality is arrived using the Green's formula and the third equality is due to the EOM of $B_{\alpha\beta}$ in the ground state (\ref{J}). Similarly, for a general $\gamma$ on the lattice, it's formed by the multiplication of contractible 't Hooft loops on a plane, whose expectation value is one similarly to $\langle T_{\gamma_{yz}}\rangle$. So $\langle T_{\gamma}\rangle$ is also one in the ground state with a general contractible loop $\gamma$.

\subsection{D. Details of family tree models}
\subsubsection{1. The [0,1,2,$D$] model}
In this section, we discuss one kind of generalized X-cube model in general dimensions which is labeled by [0,1,2,$D$]. Here the index 1 means spins are defined on links. The index 0 and $D$ means the Hamiltonian consists of the vertex term and hypercube term:
\begin{equation}
   H_{[0,1,2,D]} = -\sum_{hc} \mathcal{A}_{hc} -\sum_{x} \mathcal{B}_x^{\mu\nu}.
\end{equation}
The hypercube term is the tensor product of $\sigma_x$ on a $D$ dimensional cube. The index 2 means each vertex term is tensor product of Pauli matrices on links with two directions which is labeled by $\mu$ and $\nu$. Under OBC, the hypercube part partition function is still dual to 1D Ising chain, and the constraint of vertex terms is:
\begin{equation}
   \mathcal{B}_x^{\mu\nu}\mathcal{B}_x^{\nu\rho}\mathcal{B}_x^{\rho\mu}=\mathbb{I},
\end{equation}
 And if we dual each vertex term to the Ising interaction of two spins:
\begin{equation}
 \mathcal{B}_x^{\mu\nu}\to\sigma^{z}_{\mu}\sigma^z_\nu.
\end{equation}
Then the Hamiltonian of a single vertex is dual to the Curie-Weiss model with $D$ spins:
\begin{equation}
\hat{H}_{\text{CW}}=-\frac{1}{2D}(\sum_{j=1}^{D} \sigma^z_{j})^{2}=\frac{1}{D} \hat{H}_{\text{single vertex}}-\frac{1}{2}.
\end{equation}
The partition function in  the vertex sector is:
\begin{equation}
    \mathcal{Z}_{\textsf{vertex}}=\left(\sum_{n=0}^{D}e^{-\beta C_D^2+2\beta n(D-n)}C_D^n\right)^{N_{\textsf{vertex}}}.
\end{equation}



Thus this part of partition function is $N_{\text{vertex}}$ copies of a zero dimensional spin system. This implies there is no phase transition at finite temperature for this model.\par%
The higher-form subsystem symmetries of $[0,1,2,D]$ are all the same as that of $D=3$, the 3D X-cube model \cite{vijay2016fracton,slagle2017quantum}. As each term of the $[0,1,2,D]$ Hamiltonian contains a cube term or vertex term of the 3D X-cube model as a factor, the two anti commuting subsystem symmetries of $[0,1,2,D]$ belong to $(H^1(\mathbb{T}^2,\mathbb{Z}_2),H^1(\mathbb{T}^2,\mathbb{Z}_2))$.

\subsubsection{2. The [1,2,3,$D$] model}
In this section, we will generalize the discussion of the 4D X-cube model to general $[1,2,3,D]$ models. The Hamiltonian is:
\begin{equation}
   H_{[0,1,2,D]} = -\sum_{hc} \mathcal{A}_{hc} -\sum_{l_a,\{\perp abc\}} \hat{\mathcal{B}}_{l_a}^{\{\perp abc\}},
\end{equation}
 where $hc$ is a $D$ dimensional hypercube, $l_a$ is the link parallel to the direction $a$, and $\{\perp abc\}$ is the set of $D-3$ indices different from the mutually orthogonal $a$, $b$ and $c$ directions. The link terms live in the three dimensional leaf space expanded by $\{a,b,c\}$. The hypercube term $\mathcal{A}_{hc}$ is the tensor product of $\sigma_x$ on the plaquettes of a $D$ dimensional hupercube. The hypercube term $\hat{\mathcal{B}}_{a}^{\{\perp abc\}}$ is the tensor product of $\sigma_z$ on the four plaquettes expanded by $\{a,b\}$, or $\{a,c\}$ which  share the link $l_a$. 
\subsubsection{Field theory and non-contractible Wilson loops, ’t Hooft loops of the [1,2,3,$D$] model}
The lattice operators can be rewritten as follows :
\begin{eqnarray}
\begin{aligned} 
\hat{Z}_{i,ab}(t) & \sim \exp \left(i \int_{S}  Z_{ab}(\boldsymbol{x}, t))\right),\\
\hat{X}_{i,ab}(t)& \sim \exp \left(i \int_{\perp S} X_{ab}(\boldsymbol{x}, t)\right), \\ \hat{\mathcal{A}}_{hc}(t) & \sim \exp \left(\frac{2 \pi i}{N} \int_{hc_{D}} i^{0}(\boldsymbol{x}, t)\right) ,\\ \hat{\mathcal{B}}_{l_a}^{\{\perp abc\}}( t) & \sim \exp \left(\frac{2 \pi i}{N} \int_{hc_{D-1}} j^{0 ; a\{\perp bc\}}(\boldsymbol{x}, t)\right),
\end{aligned}
\end{eqnarray}
where $\hat{Z}_{i,ab}(t)$ and $\hat{X}_{i,ab}(t)$ are the lattice Pauli operators at time $t$, and ${\perp S}$ is the space dual to the surface $S$. $hc_{D}$ is the $D$ dimensional hypercube and $hc_{D-1}$ menas the $D-1$ dimensional hypercube dual to the link $a$. And we still demand $a\textless b$ in the indices of the gauge fields $X_{ab}$ and $Z_{ab}$ when the sum over the indices is encountered. And the densities are: 
\begin{eqnarray}
   &&i^{0}=\frac{N}{2 \pi}\left|\epsilon^{0 \alpha_{1} \alpha_{2}\cdots \alpha_{D}}\right| \left(\prod_{i=1}^{D-2}\partial_{\alpha_{i}}\right)  Z_{\alpha_{D-1} \alpha_{D}},\nonumber\\
   &&j^{0 ; \rho \sigma_{1}\cdots\sigma_{D-3}}=\frac{N}{2 \pi} \epsilon^{0 \rho\mu \sigma_{1}\cdots\sigma_{D-3} \nu} \partial_{\nu} X_{\nu \rho}.
    \end{eqnarray}
    We redefine the gauge fields:
\begin{eqnarray}
   X_{\gamma\delta} = B_{\alpha_{1}\cdots\alpha_{D-2}} |\epsilon^{0\alpha_{1}\cdots\alpha_{D-2}\gamma\delta}|,\nonumber\\ 
    Z_{\gamma\delta} = A_{\alpha_{1}\cdots\alpha_{D-2}}|\epsilon^{0\alpha_{1}\cdots\alpha_{D-2}\gamma\delta}|.
\end{eqnarray} 
    Then the densities are:
    \begin{eqnarray}\label{density2}
   &&i^{0}=\frac{N}{2 \pi} \left|\epsilon^{0 \alpha_{1} \alpha_{2}\cdots \alpha_{D}}\right| \left(\prod_{i=1}^{D-2}\partial_{\alpha_{i}}\right)  A_{\alpha_{1}\cdots\alpha_{D-2}},\nonumber\\
   &&j^{0 ; \rho \sigma_{1}\cdots\sigma_{D-3}}=\frac{N}{2 \pi}\epsilon^{0 \rho \mu\sigma_{1}\cdots\sigma_{D-3} \nu} \partial_{\nu} B_{\mu\sigma_{1}\cdots\sigma_{D-3} }.
    \end{eqnarray}\par
And the gauge transformation is given as follows:
\begin{eqnarray}\label{gaugetrans}
B_{\alpha_{1}\cdots\alpha_{D-2}}\to&& B_{\alpha_{1}\cdots\alpha_{D-2}}+\int_{x'} [B_{\alpha_{1}\cdots\alpha_{D-2}},i^{0}]\chi,\nonumber\\
=&&B_{\alpha_{1}\cdots\alpha_{D-2}}+\left(\prod_{i=1}^{D-2}\partial_{\alpha_{i}}\right)\chi,\\
A_{\alpha_{1}\cdots \alpha_{D-2}}\to&& A_{\alpha_{1}\cdots \alpha_{D-2}}+\int_{x'} [A_{\alpha_{1}\cdots \alpha_{D-2}},j^{0;\rho\sigma_1\cdots\sigma_{D-3}}]f\nonumber\\
=&&A_{\alpha_{1}\cdots \alpha_{D-2}}+\sum_{i[D-2]}\epsilon^{ \rho \alpha_{i_{1}}\cdots\alpha_{i_{D-2}}\nu}\partial_{\nu}f_{\rho\alpha_{i_{2}}\cdots\alpha_{i_{D-2}}}\nonumber,\\
\end{eqnarray}
where $\alpha_{i_{1}}\cdots\alpha_{i_{D-2}}$ is a permutation of $\alpha_{1}\cdots \alpha_{D-2}$, and the sum over $i[D-2]$ is over all possible permutations.

 We can also construct the independent 't Hooft loops and Wilson surfaces which can label ground states. For simplicity, we let the indices of  $A$ and $B$ are $x_{1}\cdots x_{D-2}$ firstly.
 
For $B$ gauge field, we define similar gauge invariant ’t Hooft loops:
\begin{eqnarray}
&&T^{i;D-1,D}(x_{1},\cdots,x_{i-1},x_{i+1},\cdots,x_{D-2})\nonumber\\
= &&\exp\left(i\int_{0}^{L_{i}a}\prod_{j\ne i} \int_{x_{j}}^{x_{j}+a} B_{x_{1}\cdots x_{D-2}}\prod_{a=1}^{D-2} d x_a\right).\nonumber\\~
 \end{eqnarray}
From the equation (\ref{density2}), we know
\begin{eqnarray}
&&j^{0,x_{D-1}x_{1}\cdots x_{i-1}x_{i+1}\cdots x_{D-2}}\propto\partial_{x_{D}}B_{x_{1}\cdots x_{i-1}x_{i}x_{i+1}\cdots x_{D-2}}-\partial_{x_{i}}B_{x_{1}\cdots x_{i-1}x_{D}x_{i+1}\cdots x_{D-2}}=0\nonumber\\\Rightarrow
&&\partial_{x_{D}}T^{i;D-1,D}\sim\int_{0}^{L_{i}a} \partial_{x_{i}}B_{x_{1}\cdots x_{i-1}x_{D}x_{i+1}\cdots x_{D-2}}d^{D-2}x=0.
\end{eqnarray}
The calculation for direction $x_{D-1}$ is similar. Thus $T$ only depends on the $D-3$ coordinates of equation above. Since there is a constraint:
\begin{eqnarray}
\prod^{L_j a}_{x_{j}=a}T^{i;D-1,D}=\prod^{L_i a}_{x_{i}=a}T^{j;D-1,D},
\end{eqnarray}
the number of independent 't Hooft loops is $\sum^{D-2}_{k=1}(-1)^{k-1}  C^k_{D-2} L^{D-2-k}$.

For $A$ gauge field, we define the following Wilson surfaces:
\begin{eqnarray}
W(x_{1},\cdots,x_{D-2})=\exp\left(i\int^{L_{D-1}a}_{x_{D-1}=0}\int^{L_{D}a}_{x_{D}=0} A_{x_{1}\cdots x_{D-2}}dx_{D-1}dx_{D}\right).\end{eqnarray}

According to the equation (\ref{density2}), we obtain $(\prod^{D-2}_{i=1}\partial_{x_{i}})\ln W=0$. Thus the independent $W$ are $W^{i}(x_{1},\cdots,x_{i-1},x_{i+1},\cdots,x_{D-2})=W(x_{1},\cdots,x_{i-1},0,x_{i+1},\cdots,x_{D-2})$. But since there is another constraint: \begin{eqnarray}
W^{i;D-1,D}(x_{1},\cdots,x_{j-1},0,x_{j+1},\cdots,x_{D-2})=W^{j;D-1,D}(x_{1},\cdots,x_{i-1},0,x_{i+1},\cdots,x_{D-2}),
\end{eqnarray}
the number of independent  't Wilson surfaces of $A$ is $\sum^{D-2}_{k=1}(-1)^{k-1}  C^k_{D-2} L^{D-2-k}$.
The commutation relations between these operators is still $\mathbb{Z}_{N}$ Heisenberg algebra,
thus there are $N^{\sum^{D-2}_{k=1}(-1)^{k-1}  C^k_{D-2} L^{D-2-k}}$ ground states of this plane.
The calculation for other planes is similar and the total ground state degeneracy is $N^{C^{2}_{D}\sum^{D-2}_{k=1}(-1)^{k-1}  C^k_{D-2} L^{D-2-k}}$.

\subsubsection{Low temperature free energy excitation expansion}
The excitation of $[1,2,3,D]$ is similar to that of [1,2,3,4] model and consists of dimension 0 fractons and dimension 2 strings. The density of fracton and dimension 2 string in the $xy$-plane are given by: 
   \begin{eqnarray}
   &&i^{0}=\delta^{D}(x),\nonumber\\
   &&j^{0 ; x \sigma_{1}\cdots\sigma_{D-3}}= \epsilon^{xy \sigma_{1}\cdots\sigma_{D-3} \sigma_{D-2} }(\delta(y+y_{1})-\delta(-y+y_{2}))\nonumber\\&&\quad\quad\quad\quad\quad\quad\quad\theta(x-x_{1})\theta(-x+x_{2})\prod_{i}\delta(\sigma_{i}) ,\nonumber\\
   &&j^{0 ; y \sigma_{1}\cdots\sigma_{D-3}}= \epsilon^{yx \sigma_{1}\cdots\sigma_{D-3} \sigma_{D-2} }(\delta(x-x_{1})-\delta(-x+x_{2}))\nonumber\\&&\quad\quad\quad\quad\quad\quad\quad\theta(y-y_{1})\theta(-y+y_{2})\prod_{i}\delta(\sigma_{i}) .
    \end{eqnarray}
When $N=2$, the energy of the string excitation is $(2D-4)L$, where $L$ is the perimeter.
Therefore, in the vicinity of $T = 0$, we expand the partition function of link terms:
\begin{equation}
     \mathcal{Z}_{\textsf{link}}(\beta) =2^{L^D}  e^{DC^{2}_{D-1}\beta L^D}(1+C^{2}_{D}L^D e^{-(8D-16)\beta}+\cdots),
\end{equation}
and the free energy is given as:
\begin{eqnarray}
    f &&= -\frac{1}{C^{2}_{D}\beta L^D}\ln \mathcal{Z}_{\textsf{link}}(\beta) \nonumber\\&&= -\frac{\ln2 }{C^{2}_{D}\beta }-\frac{1}{C^{2}_{D}\beta L^D}\left(DC^{2}_{D-1}\beta L^D+C^{2}_{D} L^D e^{-(8D-16)\beta}+\cdots \right),\nonumber\\~
\end{eqnarray}
In the thermodynamic limit $L\rightarrow \infty$, there is no singularity in the coefficient up to order $e^{-(8D-16)\beta}$. This indicates no zero temperature phase transition. In other words, there is a finite temperature phase transition.

\subsubsection{Contractible 't Hooft loops}

We can construct gauge-invariant contractible 't Hooft loops of the [1,2,3,$D$] models similar to that of the 4D X-cube model. For example, a special 't Hooft loop $T_{\gamma_{yz}}$ on the $yz$ plane is defined as follow:
\begin{equation}
\begin{aligned}
   T_{\gamma_{yz}}=\exp[i(&\int_{0}^{y_0}\int_{hc_{D-3}} B_{yu_{1}\cdots, u_{D-3}}(x_{0},y,0,u_{1},\cdots,u_{D-3})d^{D-3}udy+\int_{0}^{z_0}\int_{hc_{D-3}} B_{zu_{1}\cdots, u_{D-3}}(x_{0},y_0,z,u_{1},\cdots,u_{D-3})d^{D-3}udz\\&+\int_{y_0}^{0}\int_{hc_{D-3}} B_{yu_{1}\cdots, u_{D-3}}(x_0,y,z_0,u_{1}\cdots, u_{D-3})d^{D-3}udy+\int_{z_0}^{0}\int_{hc_{D-3}} B_{z}(x_0,0,z,u_{1}\cdots, u_{D-3})d^{D-3}udz)],
\end{aligned}
\end{equation}
where $hc_{D-3}$ is a $D-3$ dimensional cube with basis $u_{1}\cdots, u_{D-3}$. 

When $N=2$ for our series of models, the 't Hooft loop is the same as the Figure \ref{loop}.
Thus the proof of perimeter/area law is similar to that of [1,2,3,4] model. In the high temperature regime $\beta\ll 1$, the leading contribution of the denominator is the product of all the $\mathbb{I}\cosh\beta$ terms and the leading contribution to the numerator is the production of $B^{\mu}_l\sinh{\beta}$ terms inside $\Sigma$ and $\mathbb{I}\cosh\beta$ elsewhere. Here $\Sigma$ is the minimum surface whose boundary is $\gamma$. Thus $\left\langle T_{\gamma
}\right\rangle$ is:
\begin{equation}
    \begin{aligned}
    \left\langle T_{\gamma}\right\rangle & =\frac{1}{\mathcal{Z}_{\textsf{link}}} \operatorname{Tr}\left[T_{\gamma} \exp \left(\beta\sum_{l,\mu}\mathcal{B}_{l}^{\{\perp\mu\}}\right)\right] \\
    & \approx\tanh{\beta}^{S[\Sigma]}=\exp (-\ln(1/\tanh{\beta})S[\Sigma]).
    \end{aligned}
    \end{equation}
    
    In the low temperature regime $\beta\gg 1$,
    we  can also assume the creation operators of lowest energy excitation are well separated.
Thus the calculation for $\langle T_{\gamma}\rangle$ is similar to that of the 4D X-cube model:

\begin{eqnarray}
\begin{aligned}
    \langle T_{\gamma}\rangle \approx \langle T_{\gamma}\rangle_{gs}\frac{e^{-E_0\beta}e^{e^{-(8D-16)\beta}(N-2P)}}{e^{-E_0\beta}e^{e^{-(8D-16)\beta}N}} = \langle T_{\gamma}\rangle_{gs}e^{-2e^{-(8D-16)\beta} P}.
    \end{aligned}
\end{eqnarray}
Here $\langle T_{\gamma}\rangle_{gs}$ is the expectation value of $T_{\gamma}$ in the ground state. Similar to the [1,2,3,4] model, we can prove it to be one by the Green’s formula and the EOM in the ground state. Thus we show that $\langle T_{\gamma}\rangle$ obeys the perimeter/area law in the low/high temperature regime.

\end{widetext}
\end{document}